\documentclass[12pt]{article}

\usepackage{titlesec} 
\titleformat*{\section}{\large\bfseries}
\titleformat*{\subsection}{\large\bfseries}

\usepackage{hyperref} 
\usepackage[english]{babel}
\usepackage{mathtools, amssymb} 
\usepackage{mathtext} 
\usepackage{bm} 
\usepackage{amsthm} 
\usepackage{dsfont}
\usepackage{tikz}
\usetikzlibrary{patterns}

\hypersetup{colorlinks=true, linktocpage=false, linkcolor=blue!80!black, citecolor=red!85!black, urlcolor=blue} 

\def\cbk{\color{black}}

\renewcommand{\imath}{\mathrm{i}}
\renewcommand{\Re}{\operatorname{Re}}
\renewcommand{\Im}{\operatorname{Im}}
\renewcommand{\phi}{\varphi}

\usepackage{tocloft}

\numberwithin{equation}{section}

\begin{document}
\thispagestyle{empty}

\begin{center}
{\bf \large
	Reflection operator and hypergeometry II: \\[6pt]
	$SL(2,\mathbb{C})$ spin chain}
\vspace{0.7cm}

{P. Antonenko$^{\dagger\ast}$, N. Belousov$^{\dagger\circ}$, S. Derkachov$^{\dagger}$, P. Valinevich$^{\dagger}$}
\vspace{0.7cm}

{\small \it
	$^\dagger$Steklov Mathematical Institute, Fontanka 27, \\St.~Petersburg, 191023, Russia\vspace{0.3cm}\\
	$^\ast$Leonhard Euler International Mathematical Institute,\\ Pesochnaya nab. 10, St.~Petersburg, 197022, Russia\vspace{0.3cm}\\
	$^\circ$National Research University Higher School of Economics,\\ Myasnitskaya 20, Moscow, 101000, Russia}
	
\end{center}

\vspace{0.1cm}

\begin{abstract} \noindent
We consider noncompact open $SL(2, \mathbb{C})$ spin chain and construct eigenfunctions of $B$-element of monodromy matrix for the simplest case of the chain with one site. The reflection operator appearing in this construction can be used to express eigenfunction for $n$ sites in terms of the eigenfunction for $n-1$ sites, this general result is briefly announced. We prove orthogonality and completeness of constructed eigenfunctions in the case of one site, express them in terms of the hypergeometric function of the complex field and derive the equation for the reflection operator with the general $SL(2,\mathbb{C})$-invariant $\mathbb{R}$-operator.
\end{abstract}

\newpage

 \tableofcontents

 \newpage

\section{Introduction}

The closed spin chain with the symmetry group $SL(2, \mathbb{C})$ is an
integrable model with various applications in quantum field theory \cite{L93,L94,FK,DKM,DKO}. It was studied in the framework of the quantum inverse scattering method~\cite{F}, and many important objects of this method ($Q$-operators,
\mbox{$R$-operators}, separated variables representation) are known \cite{DKM,DM14}. Separation of variables (SoV) \cite{S1,S2} is performed by diagonalization of element $B(u)$ of monodromy matrix. Unitarity of the SoV transform for the closed $SL(2,\mathbb{C})$ spin chain or, equivalently, orthogonality and completeness of eigenfunctions of $B(u)$ is proven in~\cite{M}.

In this paper we consider \textit{open $SL(2,\mathbb{C})$ spin chain}. To define this model we follow Sklyanin's approach~\cite{S}. The construction of the monodromy matrix is different from the previous case and involves $K$-matrix --- a solution to the reflection equation \cite{Ch, S, KS}. Eigenfunctions of the $B$-element of monodromy matrix can be constructed recursively with a new ingredient, the  $\mathcal{K}$-operator that
solves the reflection equation with the Lax matrix of the model.
Note that in the simplest situation, when $K$-matrix and
$\mathcal{K}$-operator are identities, eigenfunctions of the $B$-element
of monodromy matrix are constructed in \cite{DMV}.

In the present paper we focus on the simplest case of spin chain with one site and
prove orthogonality and completeness of eigenfunctions. Our main computational tool is the diagram technique \cite{DKM}, which significantly simplifies proofs and makes them visual.

\section{Open $SL(2,\mathbb{C})$ spin chain}

To describe the model we recall the main formulas for the \textit{principal series representation} of the group $SL(2,\mathbb{C})$~\cite[Ch.~III]{GGV}. It is defined on the space $\mathrm{L}^2(\mathbb{C})$ of square integrable functions on $\mathbb{C}$ with the scalar product
\begin{equation} \label{sp}
	\langle \Phi|\Psi\rangle = \int \mathrm{d}^2z \; \overline{\Phi(z,\bar{z})}\,\Psi(z, \bar{z}) .
\end{equation}
The integration measure is the Lebesque measure in $\mathbb{C}$: $\mathrm{d}^2 z = \mathrm{d}\Re z \; \mathrm{d}\Im z$, and in what follows by $\int \mathrm{d}^2 z$ we denote integration over the whole complex plane.

To avoid misunderstanding, let us remark that below we denote complex conjugation of a number $a \in \mathbb{C}$ as~$a^*$, so that in generic situation $\bar{a}$ does not mean complex conjugation of~$a$. There are two traditional exceptions: for~the~variable $z$ we use~\mbox{$\bar{z} \equiv z^*$}, and complex conjugated functions are \mbox{denoted} as~$\overline{\Psi(z, \bar{z})}$. Besides, for simplicity we display only the holomorphic argument of functions~\mbox{$\Psi(z) \equiv \Psi(z, \bar{z})$}.  

The representation is parametrized by the pair of spins
\begin{equation}
	\bm{s} = (s, \bar{s}) \,, \qquad 2(s - \bar{s}) \in \mathbb{Z}.
\end{equation}
The group element
\begin{equation}
	g =
	\begin{pmatrix}
		a & b \\
		c & d
	\end{pmatrix}, \qquad ad - bc = 1,
\end{equation}
is represented by the operator $T(g)$ acting on functions $\Psi(z)$ by the formula
\begin{equation}
	[T(g) \, \Psi] (z) =
	[d-bz]^{-2s} \, \Psi \biggl(\frac{-c+az}{d-bz}\biggr).
\end{equation}
Here and in what follows for a pair $(a,\bar{a}) \in \mathbb{C}^2$ such that $a-\bar{a} \in \mathbb{Z}$ we denote the double power
\begin{align} \label{power}
	[z]^a \equiv z^a \bar{z}^{\bar{a}} = |z|^{a+\bar{a}} \, {e}^{\imath(a-\bar{a}) \arg z}.
\end{align}
The function \eqref{power} depends on both ``holomorphic'' parameter $a$ and ``antiholomorphic'' one $\bar{a}$, but for brevity we display only the ``holomorphic'' exponent. The numbers $a, \bar{a}$ are not complex conjugate in general. We impose the condition $a-\bar{a} \in \mathbb{Z}$, so that the function $[z]^a$ is single-valued in the whole complex plane. In addition, for $\rho \in \mathbb{R}$ we denote
\begin{equation}
	[z]^{\rho+a} \equiv z^{\rho+a} \bar{z}^{\rho+\bar{a}} .
\end{equation}

The principal series representation is unitary if $s$ and $\bar{s}$ satisfy the condition
\begin{equation} \label{scond}
	s^\ast + \bar{s} = 1 ,
\end{equation}
where $s^\ast$ denotes the complex conjugation of $s$. Together with the condition $2(s - \bar{s}) \in \mathbb{Z}$ this gives the parametrization of spins
\begin{equation}\label{sparam}
	s = \frac{n_s+1}{2} + \imath\nu_s, \qquad
	\bar{s} = \frac{-n_s+1}{2} + \imath\nu_s \,,
	\qquad
	n_s \in \mathbb{Z} + \sigma, \quad \nu_s \in \mathbb{R},
\end{equation}
where for the rest of the paper we fix the parameter $\sigma$
\begin{equation}
	\sigma \in \bigl\{ 0, \tfrac{1}{2} \bigr\}.
\end{equation}

Generators of the representation are defined in the standard way as the coefficients of decomposition of the map $T(g)$ in the neighbourhood of unity. They have the form
\begin{align} \label{gen}
	\begin{aligned}
		&S = z\partial_z + s, \qquad S_- = -\partial_z, \qquad
		S_+ = z^2\partial_z+2sz, \\[2pt]
		&
		\bar{S} = \bar{z}\partial_{\bar{z}} + \bar{s}, \qquad \bar{S}_- = -\partial_{\bar{z}}, \qquad
		\bar{S}_+ = \bar{z}^2\partial_{\bar{z}}+2\bar{s}\bar{z} .
	\end{aligned}
\end{align}
We call $S, S_{\pm}$ holomorphic generators and $\bar{S}, \bar{S}_\pm$ --- antiholomorphic ones. They obey the standard commutation relations of the Lie algebra $sl(2,\mathbb{C})$
\begin{align}\label{Scomm}
	[S_+, S_-] = 2S, \qquad [S,S_\pm] = \pm S_\pm,
\end{align}
and similarly for the antiholomorphic generators. Note that the holomorphic generators commute with the antiholomorphic ones. Two types of generators are conjugate to each other with respect to the scalar product~\eqref{sp}.
\begin{equation} \label{Shc}
	S^\dagger = -\bar{S}, \qquad S_-^\dagger = -\bar{S}_-, \qquad
	S_+^\dagger = -\bar{S}_+
\end{equation}
due to the assumption~\eqref{scond}.

In the theory of integrable systems an important role is played by Lax matrices. The spin chain under consideration is defined by holomorphic and antiholomorphic Lax matrices of the form
\begin{align} \label{L}
	L(u) = \begin{pmatrix}
		u + S & S_- \\
		S_+ & u - S
	\end{pmatrix},
	\qquad
	\bar{L}(\bar{u}) = \begin{pmatrix}
		\bar{u} + \bar{S} & \bar{S}_- \\
		\bar{S}_+ & \bar{u} - \bar{S}
	\end{pmatrix}.
\end{align}
Another useful notation for the Lax matrix reads
\begin{align}
	L(u) \equiv L(u + s - 1, u - s),
\end{align}
where the parameters $u_1 = u + s - 1$, $u_2 = u - s$ appear naturally in the factorized expression
\begin{align} \nonumber
	L(u_1, u_2) &
	=
	\begin{pmatrix}
		1 & 0 \\
		z & 1
	\end{pmatrix}
	\begin{pmatrix}
		u_1 & - \partial_z \\
		0 & u_2
	\end{pmatrix}
	\begin{pmatrix}
		1 & 0 \\
		-z & 1
	\end{pmatrix} \\[6pt] \label{L2}
	& = \begin{pmatrix}
		u_1 + 1 + z \partial_z & - \partial_z \\[3pt]
		z^2 \partial_z + (u_1 - u_2 + 1) z & u_2 - z \partial_z
	\end{pmatrix}.
\end{align}
For the construction of the open spin chain we also need $K$-matrices
\begin{align}\label{K}
	K(u) = \begin{pmatrix}
		\imath \alpha & u - \frac{1}{2} \\[6pt]
		\gamma^2 \bigl(u - \frac{1}{2} \bigr) & \imath \alpha
	\end{pmatrix},
	\qquad
	\bar{K}(\bar{u}) = \begin{pmatrix}
		\imath \bar{\alpha} & \bar{u} - \frac{1}{2} \\[6pt]
		\bar{\gamma}^2 \bigl(\bar{u} - \frac{1}{2} \bigr) & \imath \bar{\alpha}
	\end{pmatrix},
\end{align}
which solve the reflection equation \cite{Ch, S, KS}. The $K$-matrix considered in the $SL(2,\mathbb{R})$ case \cite[eq. (1.12)]{ABDK} can be obtained by the change $\gamma = i\beta$.

In the present work the attention is payed to the open $SL(2,\mathbb{C})$ spin chain. The Hilbert space of the chain consisting of $n$ sites is given by the direct product of $n$ copies of $\mathrm{L}^2(\mathbb{C})$, i.e. the space of square integrable functions of~$n$ variables $z_1, \ldots, z_n$. To $k$-th site one associates holomorphic Lax matrix
\begin{align}
	L_k(u) = \begin{pmatrix}
		u + z_k \partial_{z_k} + s & - \partial_{z_k} \\[4pt]
		z_k^2  \partial_{z_k} + 2s z_k & u - z_k \partial_{z_k} - s
	\end{pmatrix}
\end{align}
and the analogous antiholomorphic one $\bar{L}_k(\bar{u})$. Quantum integrals of motion of the model are constructed with the help of the monodromy matrix
\begin{equation} \label{T}
	T(u) = L_n(u) \cdots L_1(u) K(u) L_1(u) \cdots L_n(u) = \begin{pmatrix}
		A(u) & B(u) \\
		C(u) & D(u)
	\end{pmatrix} ,
\end{equation}
and its antiholomorphic counterpart $\bar{T}(\bar{u})$, which is defined by means of $\bar{L}_k(\bar{u})$ and $\bar{K}(\bar{u})$ in the same way.
The monodromy matrix satisfies the reflection equation
\begin{multline}\label{refl_eq}
R(u-v)\, \bigl(T(u)\otimes \bm{1}\bigr)\,R(u+v-1)\, \bigl(\bm{1} \otimes T(v)\bigr) \\[6pt]
= \bigl(\bm{1} \otimes T(v)\bigr)\,R(u+v-1)\,\bigl(T(u)\otimes \bm{1}\bigr)\, R(u-v).
\end{multline}
Here $R(u)$ is the Yang's $R$-matrix that acts in the tensor product $\mathbb{C}^2\otimes \mathbb{C}^2$
\begin{equation} \label{R}
	R(u) = u  + P, \qquad P \, a \otimes b = b \otimes a.
\end{equation}
From \eqref{refl_eq} and its antiholomorphic analogue we obtain the commutativity
\begin{equation}
	[B(u), B(v)] = 0, \qquad [\bar{B}(\bar{u}), \bar{B}(\bar{v})] = 0 .
\end{equation}
Note that the holomorphic and antiholomorphic operators also commute with each other
\begin{equation}
	[B(u), \bar{B}(\bar{v})] = 0.
\end{equation}
Our main objective is the simultaneous diagonalization of operators $B(u)$ and $\bar{B}(\bar{u})$. In this paper we study in detail the case $n=1$ and derive the reflection operator, which is the key element in the construction of eigenfunctions for arbitrary $n$. Most of the results and proofs are analogous to the~case of~$SL(2, \mathbb{R})$ spin chain studied in the first paper of this series~\cite{ABDK}.

\subsection{Eigenfunctions for one site} \label{sect_eigenfunc}
In this section we collect our main results concerning the spin chain with one site.
In the case $n = 1$ the operator $B(u)$ has the form
\begin{equation} \label{Bn1}
	B(u) =  \biggl(u - \frac{1}{2} \biggr) \bigl(u^2 - H^s\bigr),
\end{equation}
where $H^s$ can be written in terms of holomorphic generators~\eqref{gen}
\begin{equation}\label{H}
	\begin{aligned}
		H^{s} &= S^2 - \gamma^2 S_-^2 - 2\imath \alpha S_- \\[6pt]
		& =(z^2 - \gamma^2) \partial_z^2 + (2s + 1) z \partial_z + 2 \imath \alpha \partial_z + s^2.
	\end{aligned}
\end{equation}
One can see that the spectral problem for $B(u)$ is equivalent to the spectral problem for $H^s$. Similarly for its antiholomorphic counterpart
\begin{equation}
	\bar{B}(\bar{u}) = \biggl(\bar{u} - \frac{1}{2} \biggr) \bigl(\bar{u}^2 -\bar{H}^{\bar{s}}\bigr)
\end{equation}
and the operator
\begin{equation}\label{Hb}
	\begin{aligned}
		\bar{H}^{\bar{s}} &= \bar{S}^2 - \bar{\gamma}^2 \bar{S}_-^2 - 2\imath \bar{\alpha} \bar{S}_- \\[6pt]
		&= (\bar{z}^2 - \bar{\gamma}^2) \partial_{\bar{z}}^2 + (2\bar{s} + 1) \bar{z} \partial_{\bar{z}} + 2 \imath \bar{\alpha} \partial_{\bar{z}} + \bar{s}^2.
	\end{aligned}
\end{equation}
Let us impose the condition on these operators
\begin{equation} \label{Hcond}
	(H^s)^\dagger = \bar{H}^{\bar{s}},
\end{equation}
where the hermitian conjugation is taken with respect to the scalar product~\eqref{sp}. Operators $H^s$ and $\bar{H}^{\bar{s}}$ commute, so that under the condition \eqref{Hcond} we can construct commuting self-adjoint combinations $H^s+\bar{H}^{\bar{s}}$, $\imath (H^s-\bar{H}^{\bar{s}})$.

The holomorphic and antiholomorphic generators are conjugate to each other~\eqref{Shc}: $S^\dagger = - \bar{S}$, $S_-^\dagger = - \bar{S}_-$. Then the condition \eqref{Hcond} is satisfied if
\begin{equation}
	\bar{\alpha} = \alpha^\ast, \qquad \bar{\gamma}^2 = (\gamma^\ast)^2.
\end{equation}
Since the operator $\bar{H}^{\bar{s}}$ \eqref{Hb} depends on $\bar{\gamma}^2$, without loss of generality we may assume
\begin{equation} \label{gam_al_cond}
	\bar{\alpha} = \alpha^\ast, \qquad \bar{\gamma} = \gamma^*.
\end{equation}
Under the above assumption we look for the joint eigenfunctions of operators~$H^s$~and~$\bar{H}^{\bar{s}}$
\begin{equation} \label{H_spec_prob}
	H^s \, \Psi_{\bm{x}}(z) = x^2 \, \Psi_{\bm{x}}(z), \qquad
	\bar{H}^{\bar{s}} \, \Psi_{\bm{x}}(z) = \bar{x}^2 \, \Psi_{\bm{x}}(z) ,
\end{equation}
where we denoted the spectral parameters $\bm{x} = (x, \bar{x})$.

The joint eigenfunctions are constructed with the help of the \textit{reflection operator}~$\mathcal{K}(\bm{s},\bm{x})$ that acts on functions of $z,\bar{z}$ and satisfies two reflection equations with holomorphic Lax and $K$-matrices~\eqref{L2},~\eqref{K}
\begin{align} \label{Kdef}
	& \begin{aligned}
		&
		\mathcal{K}(\bm{s},\bm{x}) \, L(u +x - 1, u - s) \, K(u) \, L(u  + s - 1, u - x) \\[4pt]
		&
		= L(u + s - 1, u - x) \, K(u) \, L(u + x - 1, u - s) \, \mathcal{K}(\bm{s},\bm{x})
	\end{aligned}
\end{align}
and their antiholomorphic counterparts
\begin{align}
	&\begin{aligned}
		&
		\mathcal{K}(\bm{s},\bm{x}) \, \bar{L}(\bar{u} +\bar{x} - 1, \bar{u} - \bar{s}) \, \bar{K}(\bar{u}) \, \bar{L}(\bar{u}+\bar{s}-1, \bar{u} - \bar{x}) \\[4pt]
		&
		\label{Kdef_a}
		= \bar{L}(\bar{u}+\bar{s}-1, \bar{u} - \bar{x}) \, \bar{K}(\bar{u}) \, L(\bar{u} + \bar{x} - 1, \bar{u} - \bar{s}) \, \mathcal{K}(\bm{s},\bm{x}).
	\end{aligned}
\end{align}
The first relation~\eqref{Kdef} is equivalent to three independent operator equations on reflection operator, see \eqref{Krel}. One of them includes the operator $H^s$~\eqref{H}
\begin{align} \label{HK}
	H^s \, \mathcal{K}(\bm{s},\bm{x})
	= \mathcal{K}(\bm{s},\bm{x}) \, H^x.
\end{align}
Similarly, from \eqref{Kdef_a} we have
\begin{equation} \label{barHK}
	\bar{H}^{\bar{s}} \, \mathcal{K}(\bm{s},\bm{x})
	= \mathcal{K}(\bm{s},\bm{x}) \, \bar{H}^{\bar{x}} .
\end{equation}
Acting on $1$ in the last two equations we obtain
\begin{align} \label{eigenfunc}
	& H^s \, \mathcal{K}(\bm{s},\bm{x}) \cdot 1 = \mathcal{K}(\bm{s},\bm{x}) \, H^x \cdot 1 = x^2 \, \mathcal{K}(\bm{s},\bm{x}) \cdot 1, \\[6pt] \label{eigenfunc2}
	& \bar{H}^{\bar{s}} \, \mathcal{K}(\bm{s},\bm{x}) \cdot 1 = \mathcal{K}(\bm{s},\bm{x}) \, \bar{H}^{\bar{x}} \cdot 1 = \bar{x}^2 \, \mathcal{K}(\bm{s},\bm{x}) \cdot 1.
\end{align}
Thus, $\mathcal{K}(\bm{s},\bm{x})\cdot 1$ is the joint eigenfunction of the operators $H^s$ and $\bar{H}^{\bar{s}}$.

All of this can be regarded as an induction step from the eigenfunction of $B_{n=0}(u)$ to the eigenfunction of $B_{n=1}(u)$. The operator $B_{n=0}$ is an element of the matrix $K(u)$, i.e. it is a number and its eigenfunction is $1$. Acting by $\mathcal{K}(\bm{s},\bm{x})$ on $1$ we obtain an eigenfunction of $B_{n=1}(u)$, which is the function of one complex variable $z$.

Now let us formulate our main results concerning the reflection operator and eigenfunctions. It is convenient to denote
\begin{align}
	g = \frac{1}{2} + \frac{\imath \alpha}{\gamma}, \qquad \bar{g} = \frac{1}{2} + \frac{\imath \bar{\alpha}}{\bar{\gamma}},
\end{align}
and below we frequently use this pair of parameters instead of $(\alpha, \bar{\alpha})$. Apart from the conditions on the parameters \eqref{sparam}, \eqref{gam_al_cond}, in what follows we always assume \vspace{-0.3cm}
\begin{align}
	x - \bar{x} \in \mathbb{Z} + \sigma, \qquad g - \bar{g} \in \mathbb{Z} + \sigma,
\end{align}
where $\sigma \in \{0, 1/2\}$ is fixed by the spins \eqref{sparam}. The second assumption together with the conditions \eqref{gam_al_cond} gives the following parametrization
\begin{equation}
	\;\; g = \frac{n_g+1}{2} + \imath\nu_g, \qquad
	\bar{g} = \frac{-n_g+1}{2} + \imath\nu_g \,,
	\qquad
	n_g \in \mathbb{Z} + \sigma, \;\;\; \nu_g \in \mathbb{R}.
\end{equation}
Consequently, the complex conjugation rule for the pair $(g, \bar{g})$
\begin{align}\label{gprop}
	g^* + \bar{g} = 1
\end{align}
is the same as for the spins $(s,\bar{s})$ \eqref{scond}.

In Section \ref{Kder} we derive the explicit formula for the action of reflection operator $\mathcal{K}(\bm{s}, \bm{x})$ on a function $\Psi(z)$
\begin{align}\label{Kexpl-2}
	\begin{aligned}
		\bigl(\mathcal{K}(\bm{s},\bm{x}) \, \Psi \bigr)(z) &= \frac{1}{\pi \, \bm{\Gamma}(s- x) } \; [z + \gamma]^{g - s} \, [z - \gamma]^{1 - s -g} \\[6pt]
		& \times \int \mathrm{d}^2 w \; \frac{[w + \gamma]^{x - g} \, [w - \gamma]^{x + g - 1} }{  [z - w]^{x - s + 1} } \; \Psi(w).
	\end{aligned}
\end{align}
Here $\bm{\Gamma}(a)$ is the gamma function of the complex field~\cite[Section~1.4]{GGR}, \cite[Section 1.3]{N}
\begin{equation}
	\bm{\Gamma}(a) = \frac{\Gamma(a)}{\Gamma(1 - \bar{a})}.
\end{equation}
It depends on two parameters $(a, \bar{a}) \in \mathbb{C}^2$ such that $a - \bar{a} \in \mathbb{Z}$, but for brevity we display only the first one. In addition, for $\rho \in \mathbb{R}$ we write
\begin{align}
	\bm{\Gamma}(a + \rho) \equiv \frac{\Gamma(a + \rho)}{\Gamma(1 - \bar{a} - \rho)}
\end{align}
and for the products of gamma functions we use compact notation
\begin{align}
	\bm{\Gamma}(a \pm b, c,d) = \bm{\Gamma}(a + b) \, \bm{\Gamma}(a - b) \, \bm{\Gamma}(c) \, \bm{\Gamma}(d).
\end{align}
The useful properties of this function can be found in Appendix~\ref{App-DiagTech}.

Define the integral operator $[\partial_z]^a$ acting on functions $\Psi(z)$ by the formula
\begin{align}\label{der-power}
	([\partial_z]^a \, \Psi)(z) =
	\frac{1}{\pi \, \mathbf{\Gamma}(-a)} \int \mathrm{d}^2 w \,
	\frac{1}{[z-w]^{a+1}} \, \Psi(w).
\end{align}
Then the reflection operator \eqref{Kexpl-2} can be alternatively written in the form
\begin{align}\label{Kop-d}
		\mathcal{K}(\bm{s},\bm{x}) =
		[z+\gamma]^{g-s} \,
		[z-\gamma]^{1-s-g} \,
		[\partial_z]^{x-s} \,
		[z+\gamma]^{x-g} \,
		[z-\gamma]^{x+g-1}.
\end{align}
We remark that the operator~$[\partial_z]^{2\ell - 1}$ plays an important role in the representation theory of $SL(2,\mathbb{C})$, namely, it intertwines equivalent representations with spins $(\ell, \bar{\ell})$ and $(1 - \ell, 1 - \bar{\ell})$~\cite[Ch. III, Section 3.2]{GGV}. Moreover, it was used for the construction of the general $SL(2,\mathbb{C})$-invariant solution of the Yang-Baxter equation \cite[Section 2.3]{DM09}.

In addition, for the derivative operator \eqref{der-power} we have the expected relation
\begin{equation}
	[\partial_z]^a \, [\partial_z]^b = [\partial_z]^{a + b},
\end{equation}
which is equivalent to the integral identity \eqref{Chain}. Then from the formula~\eqref{Kop-d} the reflection operator clearly satisfies
\begin{align}
	\mathcal{K}(\bm{s}, \bm{x}) \, \mathcal{K}(\bm{x}, \bm{y}) = \mathcal{K}(\bm{s}, \bm{y}).
\end{align}

Section \ref{sec:psi} is devoted to the integral representations of the eigenfunctions. According to the formulas \eqref{eigenfunc}, \eqref{eigenfunc2} the eigenfunctions can be written in the form \vspace{0.15cm}
\begin{equation}\label{Psi-K}
	\Psi_{\bm{x}}(z) =  \pi \, [2 \gamma]^{s-x}  \, \bm{\Gamma}(g - x) \; \mathcal{K}(\bm{s}, \bm{x}) \cdot 1.
\end{equation}
Due to the explicit expression for the reflection operator \eqref{Kexpl-2} and reflection formula for the gamma function \eqref{gamma-refl} it is equivalent to the integral representation
\begin{align}
	\begin{aligned}
		&\Psi_{\bm{x}}(z) =  [2\gamma]^{s-x}  \, \mathbf{\Gamma}(g-x, 1 - s+x)  \\[6pt]
		&\quad \times \int \frac{\mathrm{d}^2 w}{ [z+\gamma]^{s-g} \,
		[z-\gamma]^{s+g-1} \,
		[w - z]^{x-s+1}  \,
		[w+\gamma]^{g-x} \,
		[w-\gamma]^{1-x-g} }.
	\end{aligned}
\end{align}
In Appendix \ref{App-DiagTech} we explain how such integrals with power functions can be depicted by diagrams. In this way various integral identities correspond to simple transformations of diagrams.

Using diagram technique in Section \ref{sec:psi-diag} we prove that with the above normalization the eigenfunctions have reflection symmetry
\begin{equation}
	\Psi_{\bm{x}}(z) = \Psi_{-\bm{x}}(z).
\end{equation}
Moreover, making certain diagrams transformations we derive the second integral representation for them
\begin{align}
	\begin{aligned}
		\Psi_{\bm{x}}(z) & =  [2\gamma]^{s-x}  \, \mathbf{\Gamma}(g-x, 1 - s + x)  \\[4pt]
		& \times \int \frac{\mathrm{d}^2 w}{  [w - z]^{x+s}  \, [w-\gamma]^{g-x} \,	[w+\gamma]^{1-x-g}}.
	\end{aligned}
\end{align}
With the help of this representation in Section \ref{sec:hyper} we show that the eigenfunctions are related to the hypergeometric function of the complex field \cite[\S 2]{GGR}, \cite[Section~1.6]{MN} by the formula
\begin{align}\label{Psi-2F1}
	\begin{aligned}
		\Psi_{\bm{x}}(z) & = \, \frac{ \pi \,  \mathbf{\Gamma}(g+x, g-x) }{ \mathbf{\Gamma}(s+g)}
		\\[8pt]
		&
		\times
		{}_2 F_1^{\mathbb{C}}\left[ \! \left.
		\begin{array}{c}
			s+x|\bar{s}+\bar{x}, \; s-x|\bar{s}-\bar{x} \\[3pt]
			s+g|\bar{s}+\bar{g}
		\end{array}\right| \frac{1}{2} - \frac{z}{2\gamma}\right].
	\end{aligned}
\end{align}
Last but not least in Section \ref{sec:MB} we obtain one more integral representation of Mellin-Barnes type
\begin{align}\label{PsiMB2}
	\Psi_{\bm{x}}(z) = \frac{1}{2} \int \mathcal{D}\bm{y} \;
	\frac{ [-1]^{y - s}  \; \mathbf{\Gamma}(\varepsilon-y \pm x, \, 1-s-\varepsilon+y, \, g-\varepsilon+y)}
	{[2\gamma]^{\varepsilon-y  - s } \, [z+\gamma]^{s-g} \, [z-\gamma]^{g-\varepsilon+y}  }
\end{align}
where we assume $\varepsilon \in (0,1/2)$ and that both the spectral and integration variables have the form
\begin{align}\label{xy-var}
	\begin{aligned}
		& \bm{x} = (x, \bar{x}) = \biggl( \frac{k}{2} + \imath \eta, - \frac{k}{2} + \imath \eta\biggr), &&\quad k \in \mathbb{Z} + \sigma, \quad \; \eta \in \mathbb{R}, \\[6pt]
		& \bm{y} = (y, \bar{y}) = \biggl( \frac{m}{2} + \imath \tau, - \frac{m}{2} + \imath \tau\biggr), &&\quad m \in \mathbb{Z} + \sigma , \quad \tau \in \mathbb{R}.
	\end{aligned}
\end{align}
By integration over $\bm{y}$ we mean
\begin{align}
	\int \mathcal{D} \bm{y} = \sum_{m \in \mathbb{Z} + \sigma} \, \int_\mathbb{R} \mathrm{d}\tau.
\end{align}

In Section \ref{sec:orth} we prove that eigenfunctions $\Psi_{\bm{x}}(z), \Psi_{\bm{y}}(z)$ with spectral parameters of the form~\eqref{xy-var} are orthogonal with respect to the scalar product~\eqref{sp} \vspace{-0.3cm}
\begin{align}\label{orth-2}
	\langle \Psi_{\bm{y}} | \Psi_{\bm{x}} \rangle = \mu^{-1}(\bm{x}) \, \frac{\delta(\bm{x} - \bm{y}) + \delta(\bm{x} + \bm{y})}{2}.
\end{align}
Here we denoted the delta-function with spectral parameters
\begin{align}
	\delta(\bm{x} - \bm{y}) = \delta_{km} \, \delta(\eta - \tau)
\end{align}
and the normalization coefficient
\begin{align}
	\mu(\bm{x}) = \frac{1}{4\pi^4} \Bigl| \frac{x}{\gamma} \Bigr|^2.
\end{align}
The proof of the orthogonality relies on the diagram technique described in Appendix \ref{App-DiagTech}.

With the help of the Mellin-Barnes representation \eqref{PsiMB2} in Section \ref{sec:orth} we prove the completeness relation
\begin{equation}
	\int \mathcal{D}\bm{x} \; \mu(\bm{x}) \, \Psi_{\bm{x}}(z) \, \overline{\Psi_{\bm{x}}(w)} =
	\delta^{2}(z - w),
\end{equation}
where $\delta^2(z)$ is the delta-function in the complex plane
\begin{align}
	\delta^2(z) = \delta(\Re z) \, \delta(\Im z).
\end{align}
Notice that the measure in the completeness relation $\mu(\bm{x})$ agrees with the orthogonality \eqref{orth-2}.

Finally, in Section \ref{sec:refl} we show that the operator $\mathcal{K}(\bm{s},\bm{x})$ \eqref{Kop-d} also satisfies the reflection equation
\begin{align}
	\begin{aligned}
		& \mathcal{K}_1(\bm{x}) \; \widetilde{\mathbb{R}}_{12}(\bm{x}, \bm{y}) \; \mathcal{K}_2(\bm{y}) \; {\mathbb{R}}_{12}(\bm{x},\bm{y}) \\[6pt]
		& \quad\;\; = {\mathbb{R}}_{12}(\bm{x},\bm{y}) \; \mathcal{K}_2(\bm{y}) \; \widetilde{\mathbb{R}}_{12}(\bm{x}, \bm{y}) \; \mathcal{K}_1(\bm{x})
	\end{aligned}
\end{align}
with the $\mathbb{R}$-operators defined by the formulas
\begin{align}
	&{\mathbb{R}}_{12}(\bm{x},\bm{y}) = \mathcal{P}_{12} \; [z_{1} - z_2]^{1 - s - x} \, [\partial_{z_1}]^{y - x} \, [z_{1} - z_2]^{s + y - 1}, \\[6pt]
	&\widetilde{\mathbb{R}}_{12}(\bm{x},\bm{y}) = \mathcal{P}_{12} \; [z_{1} - z_2]^{1 - 2s} \, [\partial_{z_2}]^{x - s} \, [\partial_{z_1}]^{y - s} \, [z_{1} - z_2]^{x + y - 1},
\end{align}
where $\mathcal{P}_{12}$ is the permutation operator
\begin{align}
	\mathcal{P}_{12} \, \psi(z_1, z_2) = \psi(z_2, z_1).
\end{align}
Here the lower indices mean spaces of functions of $z_1, z_2$ and in all operators we suppressed the dependence on spin $\bm{s}$. Such $\mathbb{R}$-operators represent special cases of the most general $SL(2,\mathbb{C})$-invariant solution of the Yang-Baxter equation \cite[\S 3]{DM09}.

The orthogonality and completeness of the eigenfunctions $\Psi_{\bm{x}}(z)$ are equivalent to unitarity of the transform
\begin{align}
	[T \, \Phi] (\bm{x})= \int \mathrm{d}^2 z \; \overline{\Psi_{\bm{x}}(z)} \, \Phi(z)
\end{align}
that maps the initial Hilbert space to the space of square integrable functions with respect to the measure $\mu(\bm{x}) \, \mathcal{D}\bm{x}$ and invariant under reflection~\mbox{$\bm{x} \to - \bm{x}$}. It will be interesting to establish the relation between this transform and the complex analogue of the index hypergeometric transform that also has the hypergeometric function in the kernel, but acts on different Hilbert space~\cite{MN}.

\subsection{Eigenfunctions for $n$ sites}

For the spin chain with $n$ sites operators $B(u), \bar{B}(\bar{u})$ act on functions of $n$ complex variables $z_1, \dots, z_n$. In this section we state the formula for their joint eigenfunctions
\begin{align}
	& B(u) \, \Psi_{\bm{x}_1, \dots, \, \bm{x}_n} = \left(u-\frac{1}{2}\right)
	(u^2-x_1^2) \ldots (u^2-x_n^2) \, \Psi_{\bm{x}_1, \dots, \, \bm{x}_n}  , \\[6pt]
	& \bar{B}(\bar{u}) \, \Psi_{\bm{x}_1, \dots,  \, \bm{x}_n}  = \left(\bar{u}-\frac{1}{2}\right)
	(\bar{u}^2-\bar{x}_1^2) \ldots (\bar{u}^2-\bar{x}_n^2) \, \Psi_{\bm{x}_1, \dots,  \, \bm{x}_n}
\end{align}
with pairs of spectral variables $\bm{x}_k = (x_k ,\bar{x}_k)$ such that $x_k - \bar{x}_k \in \mathbb{Z} + \sigma$.

Let us introduce the integral operator acting on functions $\Psi(z_i, z_j)$
\begin{align} \label{R_op_1}
	(\mathcal{R}_{i j}(\bm{x}) \, \Psi)(z_i, z_j) =
	\int \mathrm{d}^2 w \,
	\frac{[z_i-z_j]^{1-2s}}{[z_i-w]^{1-s+x} \, [w-z_j]^{1-s-x}} \,
	\Psi(w,z_j).
\end{align}
We note that it interchanges the arguments in two holomorphic Lax matrices
\begin{align}
	\begin{aligned}
		& \mathcal{R}_{1 2}(\bm{x}) \, L_1(u+s-1, u-s) \, L_2(u+s-1, u-x) \\[6pt]
		& = L_1(u+s-1, u-x) \, L_2(u+s-1, u-s) \, \mathcal{R}_{1 2}(\bm{x}),
	\end{aligned}
\end{align}
and the same relation holds for the antiholomorphic ones. This operator is related to the general $SL(2,\mathbb{C})$-invariant solution of the Yang-Baxter equation \cite[\S3]{DM09}.

With the help of the above $\mathcal{R}$-operator and reflection operator \eqref{Kexpl-2} we define
\begin{align}
	\begin{aligned}
		\Lambda_n(\bm{x}) & =  \mathcal{R}_{n, n-1}(\bm{x}) \, \mathcal{R}_{n-1, n-2}(\bm{x}) \ldots \mathcal{R}_{2 1}(\bm{x}) \\[4pt]
		& \times
		\mathcal{K}_1(\bm{s},\bm{x}) \,
		\mathcal{R}_{1 2}(\bm{x}) \, \mathcal{R}_{2 3}(\bm{x}) \ldots \mathcal{R}_{n-1, n}(\bm{x}).
	\end{aligned}
\end{align}
Here $\mathcal{K}_1(\bm{s},\bm{x})$ is the reflection operator acting on functions of $z_1$. Then the induction formula for the eigenfunctions reads
\begin{equation} \label{Psi_ind}
	\Psi_{\bm{x}_1, \dots, \, \bm{x}_n}(z_1, \dots, z_n) =
	\Lambda_n(\bm{x}_n) \, \Psi_{\bm{x}_1, \dots, \, \bm{x}_{n - 1}}(z_1, \dots, z_{n - 1}),
\end{equation}
or, equivalently,
\begin{equation}
	\Psi_{\bm{x}_1,  \ldots,  \, \bm{x}_n}(z_1, \dots, z_n) =
	\Lambda_n(\bm{x}_n) \, \Lambda_{n-1}(\bm{x}_{n-1}) \cdots
	\Lambda_{1}(\bm{x}_1) \cdot 1.
\end{equation}
We postpone the proof of this formula to the future work.

\section{Derivation of reflection operator} \label{Kder}

The operator $\mathcal{K}(\bm{s},\bm{x})$ acts on functions $\Psi(z) \equiv \Psi(z, \bar{z})$ and satisfies the~holomorphic reflection equation \eqref{Kdef}
\begin{align} \label{Kdef-2}
	& \begin{aligned}
		& \mathcal{K}(\bm{s},\bm{x}) \, L(u +x - 1, u - s) \, K(u) \, L(u  + s - 1, u - x) \\[4pt]
		& = L(u + s - 1, u - x) \, K(u) \, L(u + x - 1, u - s) \, \mathcal{K}(\bm{s},\bm{x})
	\end{aligned}
\end{align}
and similar antiholomorphic one \eqref{Kdef_a}. The equation \eqref{Kdef-2} is a matrix equation, the entries of matrices are operators polynomial in $u$. To obtain the system of operator equations on $\mathcal{K}(\bm{s},\bm{x})$ one needs to look at the coefficients of polynomials from both sides of relation \eqref{Kdef-2}.

Since the reflection equation \eqref{Kdef-2} is the same as in $SL(2,\mathbb{R})$ case \cite[Section~2.1]{ABDK}, we obtain the same operator equations together with their antiholomorphic counterparts
\begin{align}\label{Krel}
	\begin{aligned}
		& \mathcal{K} \, N = N \, \mathcal{K}, \qquad \mathcal{K} \, H^x = H^s \, \mathcal{K}, \qquad \mathcal{K} \, I^{x, s} = I^{s, x} \, \mathcal{K}, \\
		& \mathcal{K} \, \bar{N} = \bar{N} \, \mathcal{K}, \qquad \mathcal{K} \, \bar{H}^{\bar{x}} = \bar{H}^{\bar{s}} \, \mathcal{K}, \qquad \mathcal{K} \, \bar{I}^{\bar{x}, \bar{s}} = \bar{I}^{\bar{s}, \bar{x}} \, \mathcal{K},
	\end{aligned}
\end{align}
where the holomorphic operators are given by the formulas
\begin{align}
	&N =   (z^2-\gamma^2) \partial_z + (s+x) z, \\[6pt]
	& H^s =  (z^2-\gamma^2)\partial_z^2 + (2s+1)z\partial_z + 2\imath\alpha\partial_z + s^2, \\[6pt]
	&\begin{aligned}
		I^{s,x} &= z(z^2-\gamma^2)\partial_z^2 + (2s+1) z^2\partial_z + x(z^2-\gamma^2)\partial_z \\[4pt]
		& + 2\imath\alpha z\partial_z + s(s+x)z + N + \imath\alpha (s+x),
	\end{aligned}
\end{align}
and analogously for the antiholomorphic ones $\bar{N}, \bar{H}^{\bar{s}}, \bar{I}^{\bar{s},\bar{x}}$. To solve the above system we use the following ansatz for the reflection operator
\begin{equation} \label{K_ansatz}
	\mathcal{K}(\bm{s},\bm{x}) =
	[z+\gamma]^{g-s} \,
	[z-\gamma]^{1-s-g} \;
	W(\bm{s},\bm{x}) \;
	[z+\gamma]^{x-g} \,
	[z-\gamma]^{x+g-1},
\end{equation}
where $W(\bm{s},\bm{x})$ is an unknown operator. Here and in what follows we denote
\begin{equation} \label{g}
	g = \frac{1}{2} + \frac{\imath\alpha}{\gamma} \,, \qquad
	\bar{g} = \frac{1}{2} + \frac{\imath\bar{\alpha}}{\bar{\gamma}} \,,
\end{equation}
and notation for the double power $[z]^a$ is defined in \eqref{power}. The right hand side of \eqref{K_ansatz} is the composition of $W(\bm{s},\bm{x})$ with operators of multiplication by power functions. To have single-valued power functions we assume
\begin{align}\label{xg}
	x - \bar{x} \in \mathbb{Z} + \sigma, \qquad g - \bar{g} \in \mathbb{Z} + \sigma,
\end{align}
where $\sigma \in \{0, 1/2\}$ is fixed by the spins $(s,\bar{s})$ parametrization \eqref{sparam}. The motivation for such ansatz is explained at the end of this section.

Substituting the ansatz \eqref{K_ansatz} into the equations \eqref{Krel} we obtain the equivalent system
\begin{align} \label{Weq}
	\begin{aligned}
		&
		W \, \mathcal{N}^{x,s} = \mathcal{N}^{s,x} \, W, \qquad
		W \, \mathcal{H}^{x,s} = \mathcal{H}^{s,x} \, W, \qquad
		W \, \mathcal{I}^{x,s} = \mathcal{I}^{s,x} \, W, \\[6pt]
		&
		W \, \bar{\mathcal{N}}^{\bar{x},\bar{s}} = \bar{\mathcal{N}}^{\bar{s},\bar{x}} \, W, \qquad
		W \, \bar{\mathcal{H}}^{\bar{x},\bar{s}} = \bar{\mathcal{H}}^{\bar{s},\bar{x}} \, W, \qquad
		W \, \bar{\mathcal{I}}^{\bar{x},\bar{s}} = \bar{\mathcal{I}}^{\bar{s},\bar{x}} \, W ,
	\end{aligned}
\end{align}
where each holomorphic operator $A^{s,x} = N, H^s, I^{s,x}$ has been similarity transformed
\begin{equation}
	\mathcal{A}^{s,x} =
	[z+\gamma]^{s-g}
	[z-\gamma]^{s+g-1}
	A^{s,x} \,
	[z+\gamma]^{g-s}
	[z-\gamma]^{1-s-g},
\end{equation}
and analogously for the antiholomorphic ones. Now our aim is to solve the system \eqref{Weq} for $W(\bm{s},\bm{x})$. First, denote the generators of~$SL(2,\mathbb{C})$
\begin{equation} \label{gen_sx}
	\begin{aligned}
		& S^{s,x} = z\partial_z + \ell , \qquad
		S_-^{s,x} = -\partial_z , \qquad
		S_+^{s,x} = z^2\partial_z + 2\ell z, \\[6pt]
		& \bar{S}^{s,x} = \bar{z}\partial_{\bar{z}} + \bar{\ell}, \qquad \bar{S}^{s,x}_- = -\partial_{\bar{z}}, \qquad
		\bar{S}^{s,x}_+ = \bar{z}^2\partial_{\bar{z}}+2\bar{\ell}\bar{z} ,
	\end{aligned}
\end{equation}
with the pair of spins
\begin{equation} \label{lsx}
	\ell = \frac{1+x-s}{2} , \qquad
	\bar{\ell} = \frac{1+\bar{x}-\bar{s}}{2}.
\end{equation}
Next calculate the similarity transformed derivatives
\begin{align}
	&\begin{aligned}
		& [z+\gamma]^{s-g} \,
		[z-\gamma]^{s+g-1} \,
		\partial_z \,
		[z+\gamma]^{g-s} \,
		[z-\gamma]^{1-s-g}
		\\[6pt]
		& \;\; = \partial_z + \frac{(1-2s)z - 2\imath\alpha}{z^2-\gamma^2},
	\end{aligned} \\[10pt]
	& \begin{aligned}
		& [z+\gamma]^{s-g} \,
		[z-\gamma]^{s+g-1} \,
		\partial_z^2 \,
		[z+\gamma]^{g-s}
		[z-\gamma]^{1-s-g} \\[6pt]
		&  \;\; = \partial_z^2 + 2 \, \frac{(1-2s)z - 2\imath\alpha}{z^2-\gamma^2} \, \partial_z + \frac{(2s-1)(2s z^2 + \gamma^2)-4\alpha^2+8s\imath\alpha z}{(z^2-\gamma^2)^2}.
	\end{aligned}
\end{align}
Using the above formulas we express the operators in the system \eqref{Weq} in terms of introduced generators
\begin{align} \label{Nsx}
	\begin{aligned}
			& \mathcal{N}^{s,x} =  S_+^{s,x} + \gamma^2 S_-^{s,x} - 2\imath\alpha , \\[8pt]
		&\begin{aligned}
			\mathcal{H}^{s,x} & = -S_+^{s,x} S_-^{s,x} + (2-s-x) S^{s,x} \\[4pt]
			& + 2\imath\alpha S_-^{s,x} - \gamma^2 (S_-^{s,x})^2 + \frac{s^2+x^2-s-x}{2} ,
		\end{aligned} \\[6pt]
		& \begin{aligned}
			\mathcal{I}^{s,x} &= S_+^{s,x}S^{s,x} + \gamma^2 S^{s,x}S_-^{s,x}
			+ \frac{3-s-x}{2} \, S_+^{s,x}\\[4pt]
			& + \frac{1+s+x}{2} \, \gamma^2 S_-^{s,x}
			- 2\imath\alpha S^{s,x} - \imath\alpha.
		\end{aligned}
	\end{aligned}
\end{align}
The expressions for $\bar{\mathcal{N}}^{\bar{s},\bar{x}}$, $\bar{\mathcal{H}}^{\bar{s},\bar{x}}$, $\bar{\mathcal{I}}^{\bar{s},\bar{x}}$ are of the same form, one just needs to replace holomorphic generators and parameters with antiholomorphic ones. Therefore, the solution of the system \eqref{Weq} is given by the operator that intertwines introduced generators
\begin{align}
	W \, S^{x, s} = S^{s, x} \, W, \qquad W \, S_\pm^{x,s} = S_\pm^{s,x} \, W, \\[6pt]
	W \, \bar{S}^{\bar{x}, \bar{s}} = \bar{S}^{\bar{s}, \bar{x}} \, W, \qquad W \, \bar{S}_\pm^{\bar{x}, \bar{s}} = \bar{S}_\pm^{\bar{s}, \bar{x}}\, W.
\end{align}
Note that the interchange $s \leftrightarrows x$ corresponds to the transformation of the spin~\mbox{$\ell \to 1 - \ell$}~\eqref{lsx}. It~is known that $SL(2, \mathbb{C})$ representations with spins $(\ell, \bar{\ell})$ and $(1 - \ell, 1- \bar{\ell})$ are equivalent~\cite[Ch.~III, Section~3.2]{GGV}. The operator establishing their equivalence can be written in the form \cite[Section~2.3]{DM09}
\begin{equation}
	W = [\partial_z]^{2\ell - 1} = [\partial_z]^{x-s}.
\end{equation}
Here $[\partial_z]^a$ is an integral operator depending on two parameters $(a,\bar{a}) \in \mathbb{C}^2$ such that $a - \bar{a} \in \mathbb{Z}$. It is given by the following formula:
\begin{equation} \label{der_power}
	([\partial_z]^a \, \Psi)(z) =
	\frac{1}{ \pi  \, \mathbf{\Gamma}(-a)} \int \mathrm{d}^2 w \,
	\frac{1}{[z-w]^{a+1}} \, \Psi(w),
\end{equation}
where $\mathbf{\Gamma}(-a)$ is the gamma function of the complex field
\begin{equation}
	\bm{\Gamma}(-a) = \frac{\Gamma(-a)}{\Gamma(1 + \bar{a})}.
\end{equation}
Thus, the reflection operator admits the expression
\begin{align} \label{Kop}
	\mathcal{K}(\bm{s},\bm{x}) =
	[z+\gamma]^{g-s} \,
	[z-\gamma]^{1-s-g} \,
	[\partial_z]^{x-s} \,
	[z+\gamma]^{x-g} \,
	[z-\gamma]^{x+g-1}.
\end{align}
According to the definition \eqref{der_power}, the explicit action of this operator on $\Psi(z)$ is given by the formula
\begin{align}\label{Kexpl}
	\begin{aligned}
		\bigl(\mathcal{K}(\bm{s},\bm{x}) \, \Psi \bigr)(z) &= \frac{1}{ \pi  \, \bm{\Gamma}(s- x) } \; [z + \gamma]^{g - s} \, [z - \gamma]^{1 - s -g} \\[6pt]
		& \times \int \mathrm{d}^2 w \; \frac{[w + \gamma]^{x - g} \, [w - \gamma]^{x + g - 1} }{  [z - w]^{x - s + 1} } \; \Psi(w).
	\end{aligned}
\end{align}
Note that in Section \ref{sect_eigenfunc} we imposed the conditions $\bar{\alpha} = \alpha^*$, $\bar{\gamma} = \gamma^*$ \eqref{gam_al_cond}, which together with the assumption $g - \bar{g} \in \mathbb{Z} + \sigma$ \eqref{xg} give us the following parametrization
\begin{equation}
	g = \frac{n_g + 1}{2} + \imath \nu_g, \qquad
	\bar{g} = \frac{-n_g + 1}{2} + \imath \nu_g \,, \qquad
	n_g \in \mathbb{Z} + \sigma, \;\;  \nu_g \in \mathbb{R},
\end{equation}
similar to the parametrization of the spins $(s, \bar{s})$ \eqref{sparam}.

Now let us explain the motivation for the ansatz \eqref{K_ansatz}. In the first paper of this series we obtained the formula for the reflection operator of open $SL(2,\mathbb{R})$ spin chain \cite[Section 2.2]{ABDK}
\begin{align}\label{KR}
	\mathcal{K}(s,x) = (2 \gamma)^{s - x} \, (z + \gamma)^{g - s} \, \frac{ \Gamma\bigl( (z - \gamma) \partial_z 	+ x + g \bigr) }{ \Gamma\bigl( (z - \gamma) \partial_z 	+ s + g \bigr) } \, (z + \gamma)^{x - g}.
\end{align}
Heuristically, we can introduce the complex power of the derivative over the variable~$z$ by the following formula~\cite[eq. (2.8)]{CDKK}
\begin{equation}\label{da}
	\partial_z^a = \frac{1}{(z - \gamma)^{a}} \, \frac{ \Gamma \bigl( (z - \gamma) \partial_z + 1 \bigr)}{ \Gamma \bigl( (z - \gamma) \partial_z + 1 - a \bigr) }.
\end{equation}
This operator mimics the properties of the usual derivative, see \cite{CDKK}. Then the expression \eqref{KR} can be rewritten in the suggestive form
\begin{align}\label{KR2}
	\begin{aligned}
			\mathcal{K}(s,x) &= (2 \gamma)^{s - x} \\[6pt]
			& \times (z + \gamma)^{g - s} \, (z - \gamma)^{1 - s - g} \; \partial_z^{ x - s } \; (z - \gamma)^{x + g - 1} \,  (z + \gamma)^{x - g},
	\end{aligned}
\end{align}
which motivates the ansatz \eqref{K_ansatz}. However, in the $SL(2,\mathbb{R})$ case we consider $\gamma = \imath \beta \in \imath \mathbb{R}_{>0}$ and the Hilbert space consists of the functions analytic in the upper half-plane $\Im z > 0$. The power functions $(z - \gamma)^a$ are not single-valued in the upper half-plane, which makes the individual factors in the formula~\eqref{KR2} ill-defined.

Notice that the defining equation for the reflection operator \eqref{Kdef-2} is invariant with respect to the reflection $\gamma \to - \gamma$. Under this reflection the parameter
\begin{equation}
	g = \frac{1}{2} + \frac{\imath \alpha}{\gamma}
\end{equation}
also changes $g \to 1 - g$. It is interesting that despite the defining equation~\eqref{Kdef-2} in $SL(2,\mathbb{R})$ and $SL(2,\mathbb{C})$ cases is the same, only the $SL(2, \mathbb{C})$ solution~\eqref{Kop} is symmetric with respect to the reflection $\gamma \to - \gamma$, whereas in the $SL(2, \mathbb{R})$ case the symmetry is broken, as it can be seen from the definition \eqref{da}.

\section{Eigenfunctions}\label{sec:psi}

\subsection{Diagrammatic representations}\label{sec:psi-diag}

In this section we derive two integral representations for the one-site eigenfunctions. Both of them can be represented by certain diagrams using the technique described in Appendix \ref{App-DiagTech}.

\begin{figure}[t]
	\begin{minipage}{\textwidth}
		\centering\includegraphics[scale=0.47]{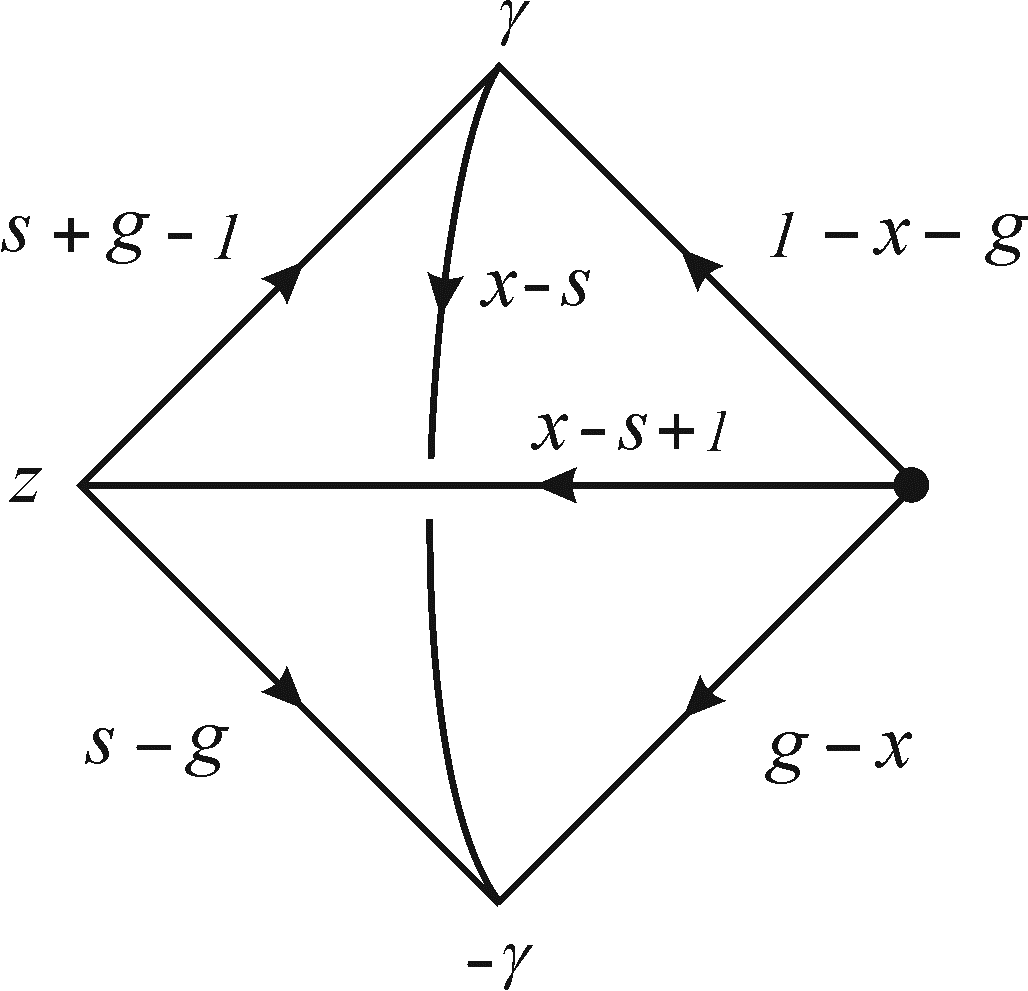}
		\caption{Diagrammatic representation of $\Psi_{\bm{x}}(z)$}
		\label{Psi_diag}
	\end{minipage}
\end{figure}

As it is explained in Section \ref{sect_eigenfunc}, the eigenfunctions can be constructed using the reflection operator \eqref{Psi-K}
\begin{align}
	\Psi_{\bm{x}}(z) =  \pi \, [2 \gamma]^{s-x}  \, \bm{\Gamma}(g - x) \; \mathcal{K}(\bm{s}, \bm{x}) \cdot 1.
\end{align}
Below we prove that with such normalization the eigenfunctions are invariant with respect to the reflection $\bm{x} \to - \bm{x}$.

The explicit action of the reflection operator is given by the formula \eqref{Kexpl}
\begin{align}
	\begin{aligned}
		\bigl(\mathcal{K}(\bm{s},\bm{x}) \, \Psi \bigr)(z) &= \frac{1}{\pi \, \bm{\Gamma}(s- x) } \; [z + \gamma]^{g - s} \, [z - \gamma]^{1 - s -g} \\[6pt]
		& \times \int \mathrm{d}^2 w \; \frac{[w + \gamma]^{x - g} \, [w - \gamma]^{x + g - 1} }{  [z - w]^{x - s + 1} } \; \Psi(w).
	\end{aligned}
\end{align}
Therefore, due to the reflection formula for the gamma function \eqref{gamma-refl}
\begin{align}
	[-1]^{s - x} = \bm{\Gamma}(s - x, 1 - s + x)
\end{align}
we obtain the first integral representation for the eigenfunctions
\begin{align}\label{Psi}
	\begin{aligned}
		&\Psi_{\bm{x}}(z) =  [2\gamma]^{s-x}  \, \mathbf{\Gamma}\left(g-x, 1 - s + x\right)  \\[6pt]
		&\quad \times \int \frac{\mathrm{d}^2 w}{ [z+\gamma]^{s-g} \,
			[z-\gamma]^{s+g-1} \,
			[w - z]^{x-s+1}  \,
			[w+\gamma]^{g-x} \,
			[w-\gamma]^{1-x-g} }.
	\end{aligned}
\end{align}
Up to gamma functions behind the integral it is depicted in Figure \ref{Psi_diag}.

\begin{figure}[t]
	\begin{minipage}{\textwidth}
		\centering\includegraphics[scale=0.47]{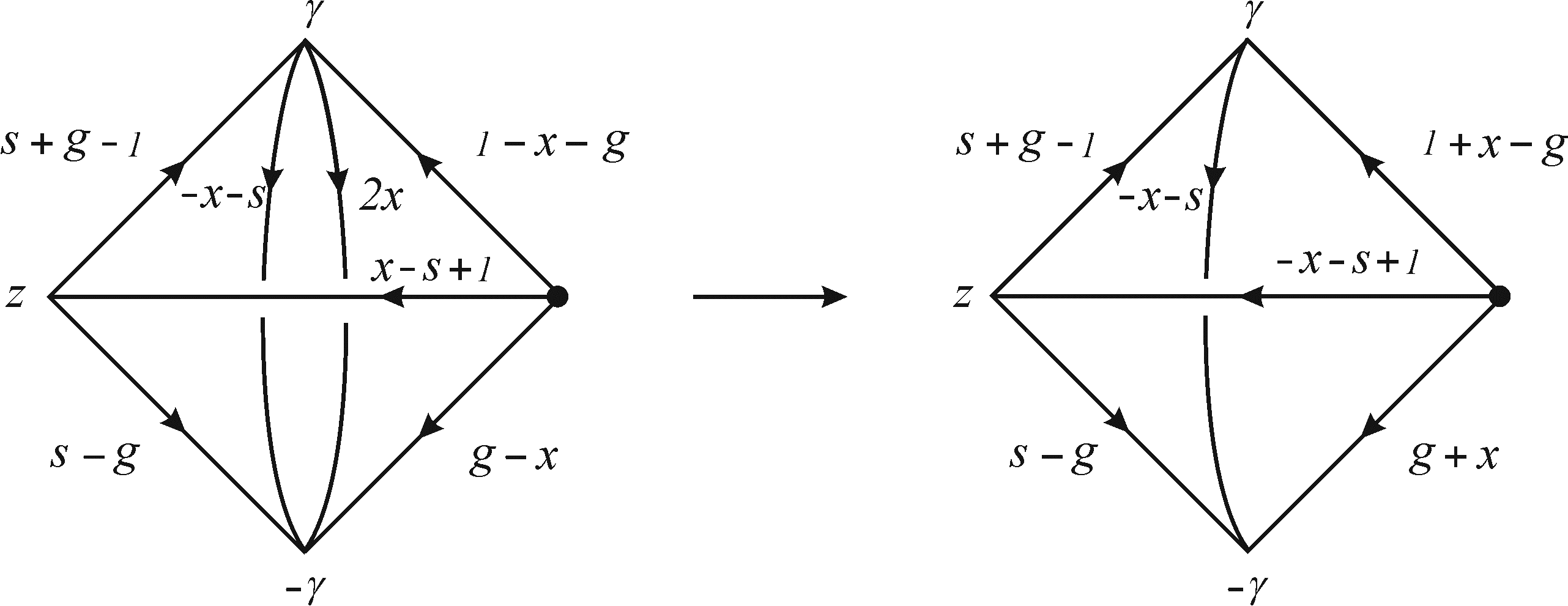}
		\caption{Reflection symmetry $\bm{x} \to - \bm{x}$}
		\label{symm_der}
	\end{minipage}
\end{figure}

Now let us prove the reflection symmetry
\begin{equation}\label{Psi_symm}
	\Psi_{\bm{x}}(z) = \Psi_{-\bm{x}}(z)
\end{equation}
using diagram technique. The left diagram in Figure~\ref{symm_der} is equivalent to the one in Figure~\ref{Psi_diag}: we just divided the vertical line corresponding to the factor~$[2\gamma]^{s-x}$ into two parts
\begin{equation}
	 [2\gamma]^{s-x} = [2\gamma]^{s + x}  \, [2\gamma]^{-2x}.
\end{equation}
Using the reduced cross relation shown in Figure~\ref{fig:Cross} we arrive at the second diagram in Figure~\ref{symm_der}. Modulo coefficient with gamma functions, this diagram depicts the function $\Psi_{-\bm{x}}(z)$. Restoring the gamma functions from \eqref{Psi} and taking into account the gamma function factor coming from the reduced cross relation \eqref{Cross_reduced} one obtains \eqref{Psi_symm}.

\begin{figure}[t]
	\begin{minipage}{\textwidth}
		\centering\includegraphics[scale=0.39]{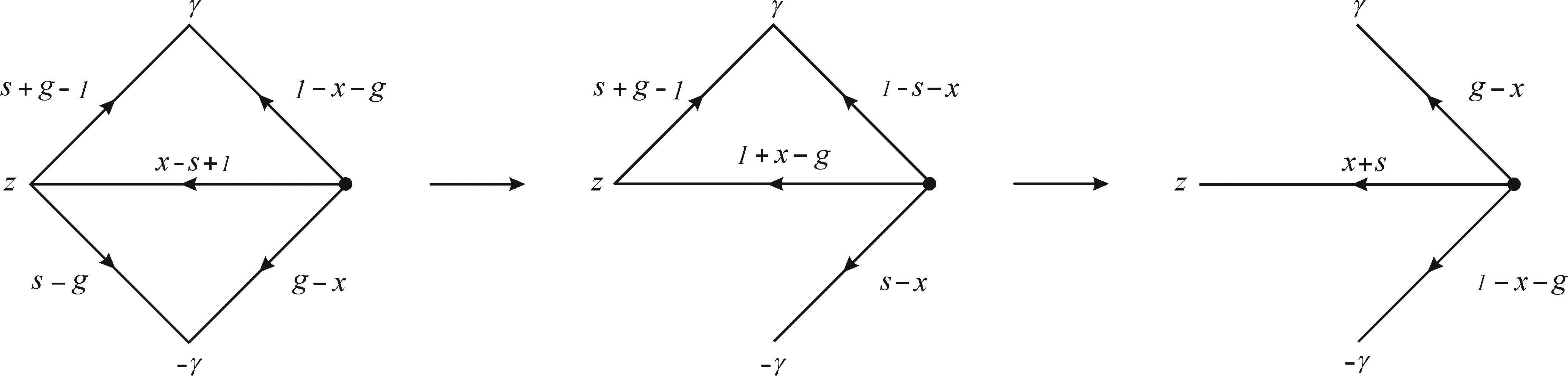}
		\caption{Derivation of the second integral representation}
		\label{psi_tripod_der}
	\end{minipage}
\end{figure}

At last, we derive the second integral representation for the eigenfunctions. Starting from the first representation \eqref{Psi} and using the same reduced cross relation \eqref{Cross_reduced} two times we arrive at
\begin{align}\label{Psi2}
	\begin{aligned}
		\Psi_{\bm{x}}(z) & =  [2\gamma]^{s-x}  \, \mathbf{\Gamma}(g-x, 1 - s + x)  \\[4pt]
		& \times \int \frac{\mathrm{d}^2 w}{  [w - z]^{x+s}  \, [w-\gamma]^{g-x} \,
			[w+\gamma]^{1-x-g}}.
	\end{aligned}
\end{align}
The corresponding diagram transformations are shown in Figure~\ref{psi_tripod_der}.

\subsection{Relation to hypergeometric function}\label{sec:hyper}
By analogy with the $SL(2,\mathbb{R})$ spin chain \cite[Section~3]{ABDK}, the one-site eigenfunction \eqref{Psi2} can be expressed in terms of the hypergeometric function. In the $SL(2,\mathbb{C})$ case it is the hypergeometric function of the complex field \cite[\S~2]{GGR}, \cite[Section~1.6]{MN}
\begin{equation}\label{2F1}
	{}_2 F_1^{\mathbb{C}}\left[ \! \left.
	\begin{array}{c}
		a|\bar{a}, \, b|\bar{b} \\
		c|\bar{c}
	\end{array}\right| v\right] = \frac{\mathbf{\Gamma}(c)}{\pi \, \mathbf{\Gamma}(b, c-b)}
	\int \mathrm{d}^2 t \; \frac{[t]^{b-1} [1-t]^{c-b-1}}{ [1-tv]^{a} },
\end{equation}
where $a-\bar{a}, \, b-\bar{b}, \, c-\bar{c} \in \mathbb{Z}$. It satisfies the pair of hypergeometric differential equations \cite[Proposition 3.7]{MN}
\begin{align}
	& \label{hg}
	\Bigl( v(1 - v)\partial_v^2 + \bigl[ c - (a+b+1) v \bigr]\partial_v -ab  \Bigr) F(v) = 0 , \\[6pt]
	& \label{hg_a}
	\Bigl( \bar{v}(1 - \bar{v})\partial_{\bar{v}}^2 + \bigl[ \bar{c} - (\bar{a}+\bar{b}+1) \bar{v} \bigr]\partial_{\bar{v}} -\bar{a}\bar{b}  \Bigr) F(v) = 0,
\end{align}
where $\bar{v} \equiv v^*$ is the complex conjugate of $v$. Notice that the spectral problem for the operator $H^s$ \eqref{H}
\begin{align}
	\Bigl( (z^2 - \gamma^2) \partial_z^2 + (2s + 1) z \partial_z + 2 \imath \alpha \partial_z + s^2 \Bigr) \Psi_{\bm{x}}(z) = x^2 \, \Psi_{\bm{x}}(z)
\end{align}
transforms into the first hypergeometric equation \eqref{hg}, if we change the variable
\begin{equation}
	v = \frac{1}{2} - \frac{z}{2\gamma}
\end{equation}
and identify the parameters
\begin{align}
	a = s+x, \qquad b = s-x, \qquad c = s+g,
\end{align}
where as before $g = 1/2 + \imath \alpha/\gamma$. Since $\gamma^* = \bar{\gamma}$ \eqref{gam_al_cond}, the same is true for the operator $\bar{H}^{\bar{s}}$ \eqref{Hb} and the second hypergeometric equation \eqref{hg_a} under identification
\begin{equation}
	\bar{a} = \bar{s}+\bar{x}, \qquad \bar{b} = \bar{s}-\bar{x}, \qquad \bar{c} = \bar{s}+\bar{g},
\end{equation}
where $\bar{g} = 1/2 + \imath \bar{\alpha}/\bar{\gamma}$.

Correspondingly, if we change the integration variable in the second integral representation of the eigenfunctions~\eqref{Psi2}
\begin{equation}
	t = -\frac{2\gamma}{w-\gamma}, \qquad
	\mathrm{d}^2 w = \frac{[2\gamma]}{[t]^2} \, \mathrm{d}^2 t,
\end{equation}
we arrive at the relation with the hypergeometric function defined above~\eqref{2F1}
\begin{align}
	\nonumber
	\Psi_{\bm{x}}(z) & = \, \frac{\pi \, \mathbf{\Gamma}(g+x, g-x)}{ \mathbf{\Gamma}(s+g)}
	\\[8pt]
	& \label{Psi_hg}
	\times
	{}_2 F_1^{\mathbb{C}}\left[ \! \left.
	\begin{array}{c}
		s+x|\bar{s}+\bar{x}, \; s-x|\bar{s}-\bar{x} \\[3pt]
		s+g|\bar{s}+\bar{g}
	\end{array}\right| \frac{1}{2} - \frac{z}{2\gamma}\right] .
\end{align}
This formula is similar to the expression for the one-site eigenfunction of the open $SL(2,\mathbb{R})$ spin chain \cite[(1.32)]{ABDK} --- parameters and the argument of the hypergeometric function have the same form. Note also that the symmetry of the hypergeometric function \cite[Proposition 3.3]{MN}
\begin{align}
	{}_2 F_1^{\mathbb{C}} \biggl[ \!
	\begin{array}{c}
	a | \bar{a}, \, b | \bar{b} \\
	c | \bar{c}
	\end{array}  \bigg| \, v \biggr] = {}_2 F_1^{\mathbb{C}} \biggl[  \!
	\begin{array}{c}
	b | \bar{b}, \, a | \bar{a} \\
	c | \bar{c}
	\end{array}  \bigg| \, v \biggr]
\end{align}
corresponds to the reflection symmetry $\bm{x} \to - \bm{x}$ of the eigenfunctions.

\subsection{Mellin-Barnes representation}\label{sec:MB}

Denote for brevity the products of gamma functions
\begin{align}
	\bm{\Gamma}(a \pm b, c, d) = \bm{\Gamma}(a + b) \, \bm{\Gamma}(a - b) \, \bm{\Gamma}(c) \, \bm{\Gamma}(d).
\end{align}
In this section we obtain one more representation for the eigenfunctions
\begin{equation} \label{PsiMB}
	\Psi_{\bm{x}}(z) = \frac{1}{2} \int \mathcal{D}\bm{y} \;
	\frac{ [-1]^{y - s}  \; \mathbf{\Gamma}(\varepsilon- y \pm x, \, 1-s-\varepsilon+y, \, g-\varepsilon+y)}
	{[2\gamma]^{\varepsilon-y  - s } \, [z+\gamma]^{s-g} \, [z-\gamma]^{g-\varepsilon+y}  }
\end{equation}
with arbitrary $ \varepsilon \in (0,1/2)$. Here we assume that both the spectral and integration variable have the form
\begin{align}\label{xparam}
	& \bm{x} = (x, \bar{x}) = \biggl( \frac{k}{2} + \imath \eta, -\frac{k}{2} + \imath \eta \biggr), && k \in \mathbb{Z}  + \sigma , \quad \; \eta \in \mathbb{R}, \\[6pt]
	& \bm{y} = (y, \bar{y}) = \biggl( \frac{m}{2} + \imath \tau, -\frac{m}{2} + \imath \tau \biggr), &&  m \in \mathbb{Z}  + \sigma , \quad \tau \in \mathbb{R}
\end{align}
and by integration over $\bm{y}$ we mean
\begin{align}
	\int \mathcal{D} \bm{y} = \sum_{m \in  \mathbb{Z} + \sigma } \,  \int_\mathbb{R} \mathrm{d}\tau.
\end{align}
Recall that the parameter $\sigma \in \{0, 1/2\}$ is fixed from the very beginning by the spins $(s, \bar{s})$ \eqref{sparam}. Notice that the reflection symmetry $\bm{x} \to - \bm{x}$ is evident from the above representation.

For the proof we need the formula \cite[eq. (3.2)]{N}
\begin{align}\label{powerMB}
	\int \mathcal{D}\bm{y} \; \bm{\Gamma}(a + y, b - y) \, [z]^{- y - a} = 2\pi \, \bm{\Gamma}(a + b) \, \frac{1}{[1 + z]^{a + b}},
\end{align}
where it is assumed that $a - \bar{a}, \, b - \bar{b} \in \mathbb{Z} + \sigma$ and
\begin{align}\label{Reab}
	\Re ( a + \bar{a}) > 0, \qquad \Re(b + \bar{b}) > 0,
\end{align}
so the contour $\tau \in \mathbb{R}$ separates series of poles from gamma functions.

To obtain the Mellin-Barnes representation \eqref{PsiMB} we start with the formula for the eigenfunction \eqref{Psi}
\begin{align}\label{Psi-3}
	\begin{aligned}
		&\Psi_{\bm{x}}(z) = [2\gamma]^{s-x} \, \mathbf{\Gamma}(g-x, 1 - s + x)  \\[6pt]
		& \quad \times \int \frac{\mathrm{d}^2 w}{[z+\gamma]^{s-g} \,
			[z-\gamma]^{s+g-1} \,
			[w - z]^{x-s+1} \,
			[w+\gamma]^{g-x} \,
			[w-\gamma]^{1-x-g}}.
	\end{aligned}
\end{align}
First, change the variable $z \to -(z - \gamma)/(w-\gamma)$ in the identity~\eqref{powerMB} and with its help rewrite the power function
\begin{align}
	\begin{aligned}
		\frac{1}{[w - z]^{x-s+1}} &=
		\frac{1}{2\pi \, \mathbf{\Gamma}(x-s+1)} \\[6pt]
		& \times \int \mathcal{D}\bm{y} \;
		\frac{[-1]^{y - s} \, \mathbf{\Gamma}(1 - s - \varepsilon + y, \, x + \varepsilon-y)}{[z-\gamma]^{1 - s - \varepsilon + y} \,
			[w-\gamma]^{x + \varepsilon-y}}.
	\end{aligned}
\end{align}
Here we assume $\varepsilon \in (0, 1/2)$ in accordance with \eqref{Reab} and the parametrizations of $\bm{s}, \bm{x}$ \eqref{sparam}, \eqref{xparam}. Then from \eqref{Psi-3} we obtain the multiple integral
\begin{align}
	\begin{aligned}
		&\Psi_{\bm{x}}(z) = \frac{[2\gamma]^{s - x} \, \bm{\Gamma}(g - x)}{2\pi}  \\[6pt]
		& \; \times \int \mathrm{d}^2w \; \mathcal{D}\bm{y} \; \frac{[-1]^{y - s} \, \mathbf{\Gamma}(1 - s - \varepsilon + y, \, x + \varepsilon-y)}{[z + \gamma]^{s - g} \, [z - \gamma]^{g + y - \varepsilon} \, [w + \gamma]^{g - x} \, [w - \gamma]^{1 + \varepsilon - g - y}}.
	\end{aligned}
\end{align}
Integrating over $w$ using the chain relation \eqref{Chain}
\begin{multline}
	\int \mathrm{d}^2 w \; \frac{1}{[w + \gamma]^{g - x} \, [w - \gamma]^{1 + \varepsilon  - g - y}} \\[6pt]
	= \frac{\pi \, [-1]^{g - x}}{ \bm{\Gamma}( 1 + \varepsilon - g - y, \, g - x, \, 1 - \varepsilon + x + y ) } \; \frac{1}{[2\gamma]^{\varepsilon - x - y}}
\end{multline}
we arrive at the desired formula \eqref{PsiMB}.

\section{Orthogonality}\label{sec:orth}

Recall the scalar product \eqref{sp}
\begin{align}
	\langle \Phi | \Psi \rangle = \int \mathrm{d}^2 z \, \overline{\Phi(z)} \, \Psi(z),
\end{align}
where integration is performed over the whole complex plane. In this section we prove that the eigenfunctions $\Psi_{\bm{x}}(z)$ are orthogonal with respect to this scalar product
\begin{equation} \label{orth}
	\langle\Psi_{\bm{y}}|\Psi_{\bm{x}}\rangle =
	2 \pi ^4 \, \Bigl| \frac{\gamma}{x} \Bigr|^2
	\bigl( \delta(\bm{x}-\bm{y}) + \delta(\bm{x}+\bm{y}) \bigr),
\end{equation}
where the spectral parameters are of the type
\begin{equation}\label{xy}
	\begin{aligned}
		& \bm{x} = (x, \bar{x}) = \biggl( \frac{k}{2} + \imath \eta, - \frac{k}{2} + \imath \eta \biggr), && \quad  k \in \mathbb{Z} + \sigma, \quad \; \eta \in \mathbb{R}, \\[6pt]
		& \bm{y} = (y, \bar{y}) = \biggl( \frac{m}{2} + \imath \tau, - \frac{m}{2} + \imath \tau \biggr), &&\quad m \in \mathbb{Z} + \sigma, \quad \tau \in \mathbb{R}
	\end{aligned}
\end{equation}
and for brevity we denote
\begin{equation}
	\delta(\bm{x}-\bm{y}) = \delta_{km} \, \delta(\eta-\tau).
\end{equation}
Note that the eigenfunctions $\Psi_{\bm{x}}$ possess the symmetry~$\bm{x} \to -\bm{x}$ (see Section \ref{sec:psi-diag}). Therefore, the scalar product also must be invariant under reflection~$\bm{x} \to -\bm{x}$. So, it is sufficient to prove the formula \eqref{orth} in the sector~$\bm{x}\neq-\bm{y}$, where the second delta-function $\delta(\bm{x}+\bm{y})$ disappears, and at the end restore full expression by symmetry.

\begin{figure}[t]
	\begin{minipage}{\textwidth}
		\centering\includegraphics[scale=0.15]{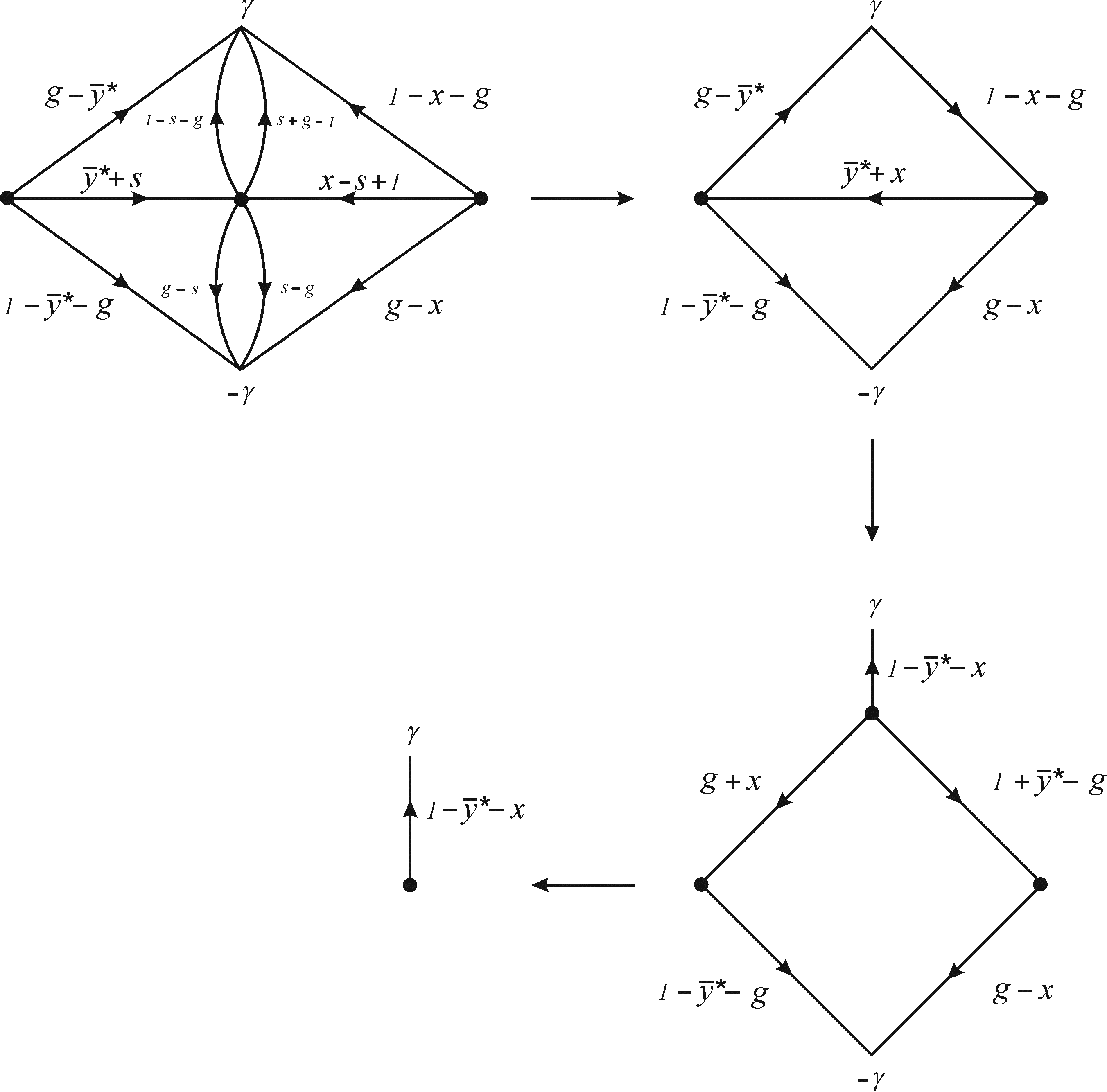}
		\caption{Derivation of the orthogonality relation}
		\label{orth_der}
	\end{minipage}
\end{figure}

For the proof we use the integral representation \eqref{Psi}
\begin{align}
	\begin{aligned}
		& \Psi_{\bm{x}}(z) = [2\gamma]^{s-x} \, \mathbf{\Gamma}\left(g-x, 1 - s + x\right)  \\[4pt]
		& \quad \times \int \frac{\mathrm{d}^2 w}{[z+\gamma]^{s-g} \,
			[z-\gamma]^{s+g-1} \,
			[w - z]^{x-s+1} \,
			[w+\gamma]^{g-x} \,
			[w-\gamma]^{1-x-g}}
	\end{aligned}
\end{align}
and diagram technique explained in Appendix \ref{App-DiagTech}. Using the rule for complex conjugation of double powers
\begin{equation} \label{power_conj}
	([z]^a)^\ast = [z]^{\bar{a}^\ast} = z^{\bar{a}^\ast} \bar{z}^{a^\ast}
\end{equation}
and conditions on the parameters \eqref{scond}, \eqref{gprop}
\begin{align}
	s^* + \bar{s} = 1, \qquad g^* + \bar{g} = 1
\end{align}
one obtains the diagrammatic representation for the scalar product $\langle\Psi_{\bm{y}}|\Psi_{\bm{x}}\rangle$ in the top left corner in Figure~\ref{orth_der} (modulo coefficient behind the integral).

The~first transformation in Figure~\ref{orth_der} is the cancellation of lines connecting the central vertex with $\gamma$ and $-\gamma$ and the use of chain relation \eqref{Chain} depicted in Figure~\ref{Rules}. On the next step we use the star-triangle relation \eqref{Star} also shown in Figure~\ref{Rules}. In the last transformation we again use two chain relations and arrive at the lower left corner:
\begin{equation} \label{PsiyPsix}
	\int \mathrm{d}^2w \, [w-\gamma]^{-1+\bar{y}^\ast+x} =
	\int \mathrm{d}^2w \, [w]^{-1+\bar{y}^\ast+x}.
\end{equation}
For $\bm{x}$ and $\bm{y}$ of the form \eqref{xy} the integral \eqref{PsiyPsix} can be easily calculated in polar coordinates $w = r e^{i\varphi}$, which gives
\begin{equation}
	\int \mathrm{d}^2w \, [w]^{-1+\bar{y}^\ast+x} =
	2\pi^2 \, \delta_{km} \, \delta(\eta-\tau) = 2\pi^2 \, \delta(\bm{x} - \bm{y}).
\end{equation}
Taking into account the factors with gamma functions appearing during diagram transformations and using the reflection formula \eqref{gamma-refl}
\begin{equation} \label{Gamma_refl}
	\mathbf{\Gamma}(a) \, \mathbf{\Gamma}(1-a) = [-1]^a = (-1)^{a-\bar{a}}
\end{equation}
we obtain the expression for the scalar product
\begin{equation} \label{PsiyPsix_0}
	\langle\Psi_{\bm{y}}|\Psi_{\bm{x}}\rangle =
	\frac{2\pi^4 \, [-1]^{x + y} \, [2\gamma] }{\mathbf{\Gamma}(1+x+y) \, \mathbf{\Gamma}(1-x-y)} \, \delta(\bm{x} - \bm{y}).
\end{equation}
The complex gamma function satisfies the difference equation \eqref{gamma-diff}
\begin{equation} \label{Gamma_recur}
	\mathbf{\Gamma}(a+1) = -a\bar{a} \, \mathbf{\Gamma}(a).
\end{equation}
Using it together with the reflection formula \eqref{Gamma_refl} we rewrite \eqref{PsiyPsix_0} as
\begin{align} \label{PsiyPsix_1}
	\langle\Psi_{\bm{y}}|\Psi_{\bm{x}}\rangle &
	= -\frac{2\pi^4 \, [2\gamma]\cbk}{(x+y)(\bar{x}+\bar{y})} \, \delta(\bm{x}-\bm{y}) \\[6pt]
	\label{PsiyPsix_2}
	& =  2 \pi^4 \, \Bigl| \frac{ \gamma }{x} \Bigr|^2  \, \delta(\bm{x}-\bm{y}).
\end{align}
Derivation of \eqref{PsiyPsix_2} is based on the tacit assumption $\bm{x} \neq \bm{y}$, only in this case all steps in Figure \ref{orth_der} have sense. This can be clearly seen from the expression in front of delta-function in \eqref{PsiyPsix_1}. This expression collects all factors appearing in intermediate steps of calculation and it is evidently singular at the point $\bm{x} = -\bm{y}$.

Removing the condition $\bm{x} \neq -\bm{y}$ and restoring the whole answer
from~\eqref{PsiyPsix_2} by symmetry we obtain the orthogonality relation~\eqref{orth}.

\section{Completeness}\label{sec:compl}

In this section we prove the completeness relation
\begin{equation} \label{compl}
	\int \mathcal{D}\bm{x} \; \mu(\bm{x}) \, \Psi_{\bm{x}}(z) \, \overline{\Psi_{\bm{x}}(w)} =
	\delta^{2}(z - w),
\end{equation}
where as before we denote
\begin{equation} \label{int_param}
	\bm{x} = (x,\bar{x}) =
 \biggl(\frac{k}{2} + \imath\eta, -\frac{k}{2} + \imath\eta \biggr), \qquad
	\int \mathcal{D}\bm{x} = \sum\limits_{k \in \mathbb{Z} + \sigma } \, \int_{\mathbb{R}} \mathrm{d}\eta
\end{equation}
and $\delta^{2}(z )$ is the two-dimensional delta function in the complex plane
\begin{equation}
	\delta^{2}(z ) = \delta(\Re z ) \, \delta(\Im z ).
\end{equation}
The integration measure
\begin{equation} \label{measure}
	\mu(\bm{x}) = \frac{1}{4\pi^4} \, \Bigl| \frac{x}{\gamma} \Bigr|^2
\end{equation}
follows from the orthogonality relation \eqref{orth} and
it is useful to rewrite it in equivalent form
\begin{equation} \label{measure1}
	\mu(\bm{x}) = \frac{ \bm{\Gamma} (s \pm x, \, 1 - s \pm x ) }{4\pi^4\, [2\gamma] \, \mathbf{\Gamma}(\pm 2x)}
\end{equation}
with the help of difference equation \eqref{gamma-diff} and reflection formula \eqref{gamma-refl} for the gamma function.

To prove the completeness relation \eqref{compl} we use Mellin-Barnes representation for the eigenfunctions \eqref{PsiMB}
\begin{equation} \label{PsiMB-2}
	\Psi_{\bm{x}}(z) = \frac{1}{2}\,\int \mathcal{D}\bm{y} \;
	\frac{[-1]^{y - s} \, \mathbf{\Gamma}(\varepsilon-y \pm x, \, 1-s-\varepsilon+y, \, g-\varepsilon+y)}
	{[2\gamma]^{\varepsilon-y - s} \, [z+\gamma]^{s-g} \, [z-\gamma]^{g-\varepsilon+y}  },
\end{equation}
where $\varepsilon \in (0,1/2)$ and the integration variable is of the type
\begin{align}
	\bm{y} = (y, \bar{y}) = \biggl( \frac{m}{2} + \imath \tau, - \frac{m}{2} + \imath \tau \biggr), \qquad m \in \mathbb{Z} + \sigma, \quad \tau \in \mathbb{R}.
\end{align}
In what follows we treat $\varepsilon$ as the regularization parameter and at the end of the calculation take the limit $\varepsilon \to 0_+$.

To write down the complex conjugated eigenfunction $\overline{\Psi_{\bm{x}}(w)}$ recall the conditions on the parameters \eqref{scond}, \eqref{gprop}
\begin{align}
	s^* + \bar{s} = 1, \qquad g^* + \bar{g} = 1
\end{align}
and how the double powers \eqref{power} are conjugated
\begin{align}
	([z]^a)^* = [z]^{\bar{a}^*}.
\end{align}
Note also that assuming $a,\bar{a}$ are of the form
\begin{equation}
	a = \frac{n}{2} + \imath \nu, \qquad
	\bar{a} = -\frac{n}{2} + \imath \nu,
	\qquad n \in \mathbb{Z}, \quad \nu \in \mathbb{R}
\end{equation}
from the definition of the gamma function \eqref{gamma-def} and the formula \eqref{gamma-swap} we have complex conjugation rule
\begin{equation}
	\mathbf{\Gamma}(\rho+a)^\ast = \bm{\Gamma}(\rho - \bar{a}) = [-1]^a \, \mathbf{\Gamma}(\rho-a),
	\qquad \rho \in \mathbb{R}.
\end{equation}
Due to all of the above remarks we have
\begin{align} \label{PsiMB1}
\overline{\Psi_{\bm{x}}(w)} =  \frac{1}{2}\,\int \mathcal{D}\bm{y}' \,
	\frac{[-1]^{y' - g} \, \mathbf{\Gamma}(\varepsilon+y'\pm x, \, s-\varepsilon-y', \, 1-g-\varepsilon-y')}{[2\gamma]^{\varepsilon+y' + s - 1} \, [w+\gamma]^{g-s} \,  [w-\gamma]^{1-g-\varepsilon-y'}
} ,
\end{align}
where integration variable is again of the type
\begin{equation}
	\bm{y}' = (y', \bar{y}') = \biggl( \frac{m'}{2} + \imath \tau', - \frac{m'}{2} + \imath \tau' \biggr), \qquad m' \in \mathbb{Z} + \sigma, \quad \tau' \in \mathbb{R}.
\end{equation}
Substituting the explicit formulas \eqref{measure1}, \eqref{PsiMB-2} and \eqref{PsiMB1}
into the left hand side of the completeness relation \eqref{compl} and taking limit $\varepsilon \to 0_+$ one obtains the multiple integral
\begin{align} \label{compl1}
	\begin{aligned}
		&
		\int \mathcal{D}\bm{x} \; \mu(\bm{x}) \, \Psi_{\bm{x}}(z) \, \overline{\Psi_{\bm{x}}(w)} \\[6pt]
		&\quad = \frac{1}{16 \pi^4} \; \lim\limits_{\varepsilon \to 0_+} \,
		\int \mathcal{D}\bm{y} \;
		\frac{ [-1]^{y - s} \; \mathbf{\Gamma}(1-s-\varepsilon+y, \, g-\varepsilon+y) }{ [z+\gamma]^{s-g} \, [z-\gamma]^{g-\varepsilon+y} }
		\\[10pt]
		&  \quad
		\times
		\int \mathcal{D}\bm{y}' \;
		\frac{ [-1]^{y' - g} \; \mathbf{\Gamma}(s-\varepsilon-y', \, 1-g - \varepsilon-y') }{  [2\gamma]^{2\varepsilon-y+y'}   \, [w+\gamma]^{g-s} \, [w-\gamma]^{1-\varepsilon-g-y'} }
		\\[10pt]
		& \quad
		\times
		\int \mathcal{D}\bm{x} \;
		\frac{ \mathbf{\Gamma}( s \pm x, \,1 - s - 4\varepsilon \pm x, \, \varepsilon-y\pm x, \, \varepsilon+y'\pm x ) }{ \mathbf{\Gamma}(\pm 2x) },
	\end{aligned}
\end{align}
where in the last $\bm{x}$-integral we introduced additional regularization
\begin{equation}
	\bm{\Gamma} ( 1 - s \pm x ) =
	\lim\limits_{\varepsilon\to 0_+} \bm{\Gamma} ( 1 - s - 4\varepsilon \pm x ).
\end{equation}
In order to integrate over $\bm{x}$ we need the complex analog of $\mathrm{sp}(1)$ Gustafson integral~\cite[eq. (2.3b)]{DM20},~\cite[eq. (65)]{SS}
\begin{equation} \label{Gust}
	\int \mathcal{D}\bm{x} \; \frac{\prod\limits_{j=1}^4 \bm{\Gamma}(a_j\pm x)}{\mathbf{\Gamma}(\pm 2x)}
	= \frac{4\pi \prod\limits_{1\leq j<k\leq 4} \mathbf{\Gamma}(a_j+a_k)}{\mathbf{\Gamma}(a_1 + a_2 + a_3 + a_4)},
\end{equation}
where the integration variable is of the same type as before \eqref{int_param} and the parameters $a_j, \bar{a}_j$ are of the form
\begin{align}\label{aj-param}
	a_j = \frac{n_j}{2} + \nu_j, \qquad \bar{a}_j = -\frac{n_j}{2} + \nu_j,
	\qquad n_j \in \mathbb{Z} + \sigma, \quad \nu_j \in \mathbb{C}.
\end{align}
The identity \eqref{Gust} holds if the contour of integration $ \eta \in \mathbb{R}$ separates the series of poles \cite{DM20}
\begin{equation}
	\imath\eta = \frac{|k-n_j| }{2} + \nu_j + p, \qquad
	\imath\eta = -\frac{|k+n_j|}{2} - \nu_j - p, \qquad
	p \in \mathbb{Z}_{\geq 0}
\end{equation}
and $\nu_j$ obey the following inequality
\begin{equation}
	\Re(\nu_1 + \nu_2 + \nu_3 + \nu_4) < 1 .
\end{equation}
In our case this condition is satisfied since in the $\bm{x}$-integral \eqref{compl1} we have
\begin{align}
		a_1 =s, \quad
		a_2 = 1 - s - 4\varepsilon, \quad
		a_3 = \varepsilon - y, \quad
		a_4 = \varepsilon + y',
\end{align}
so that according to \eqref{aj-param}
\begin{align}
		\nu_1 = \frac{1}{2} + \imath \nu_s, \quad
		\nu_2 = \frac{1}{2} - \imath \nu_s - 4\varepsilon, \quad
		\nu_3 = \varepsilon - \imath\tau, \quad
		\nu_4 = \varepsilon + \imath\tau',
\end{align}
and the needed inequality holds for small enough $\varepsilon > 0$
\begin{equation}
\Re(\nu_1 + \nu_2 + \nu_3 + \nu_4)  = 1-2\varepsilon < 1.
\end{equation}
Hence, integrating over $\bm{x}$ with the help of the Gustafson integral \eqref{Gust} and using reflection formula for the gamma functions \eqref{gamma-refl} we obtain
\begin{align}\label{x_int}
	\begin{aligned}
		& \int \mathcal{D}\bm{x} \;
		\frac{ \mathbf{\Gamma}(  s \pm x, \, 1 - s - 4\varepsilon \pm x, \, \varepsilon-y\pm x, \, \varepsilon+y'\pm x \bigr) }{ \mathbf{\Gamma}(\pm 2x) } \\[4pt]
		&\quad = 4\pi \, [-1]^{y-y'} \,  \mathbf{\Gamma} \bigl(1-4\varepsilon, \, 2\varepsilon \pm (y-y') \bigr)\\[6pt]
		&\quad \times \mathbf{\Gamma}(s + \varepsilon - y, \, s + \varepsilon+y', \, 1 - s - 3\varepsilon - y, \, 1- s - 3\varepsilon + y'  ) .
	\end{aligned}
\end{align}
The next step is to rewrite three gamma functions from the second line
\begin{multline}\label{GG}
\mathbf{\Gamma} \bigl(1-4\varepsilon, \, 2\varepsilon \pm (y-y') \bigr)  \\[6pt]
= \frac{4\varepsilon }{ (2\varepsilon+y-y')(2\varepsilon-y+y') }\,
\frac{\Gamma(1-4\varepsilon)}{\Gamma(1+4\varepsilon)} \,
\mathbf{\Gamma}\bigl(2\varepsilon+\Delta \pm (y-y')\bigr)
\end{multline}
using the pair of variables $\Delta = 1$, $\bar{\Delta} = 0$. Let us prove that the first fraction in \eqref{GG} produces the delta-function
\begin{align}\label{delta_lam}
	\lim_{\varepsilon \to 0_+} \frac{ 4\varepsilon }{ (2\varepsilon+y-y') (2\varepsilon-y+y') } = 2\pi\,\delta_{m\, m'}\,
	\delta(\tau-\tau') = 2\pi \, \delta(\bm{y}-\bm{y}').
\end{align}
First, we decompose it into the sum
\begin{align}\nonumber
		& \frac{ 4\varepsilon }{ ( 2\varepsilon+y-y') (2\varepsilon-y+y') } =
		\frac{1}{2\varepsilon+y-y'} +
		\frac{1}{2\varepsilon-y+y'} \\[6pt]
		& \label{delta_seq}
		\quad =
		-\frac{\imath}{-\imath\frac{m-m'}{2}+\tau-\tau'-2\imath\varepsilon}
		+\frac{\imath}{-\imath\frac{m-m'}{2}+\tau-\tau'+2\imath\varepsilon}.
\end{align}
If $m\neq m'$ then in the limit $\varepsilon \to 0_+$ both summands in \eqref{delta_seq} do not contain singularities in $\tau, \tau' \in \mathbb{R}$ and cancel each other. In the remaining case $m=m'$ we use Sokhotski–Plemelj theorem
\begin{equation}
	\lim\limits_{\varepsilon\to 0_+} \left(\frac{1}{\tau-\imath\varepsilon} - \frac{1}{\tau+\imath\varepsilon}\right) = 2\pi\imath\,\delta(\tau).
\end{equation}
Hence, the result for both cases is given by the formula \eqref{delta_lam}.

The rest part of the right hand side in \eqref{GG} does not contain poles on the contours of integration over $\tau$ and $\tau'$ in the limit $\varepsilon \to 0_+$, as well as the remaining gamma functions in the right hand side of~\eqref{x_int} and in~\eqref{compl1}.

Therefore, putting~\eqref{x_int},~\eqref{GG},~\eqref{delta_lam} together, integrating the delta-function and using reflection formula for the gamma functions \eqref{gamma-refl} we derive from~\eqref{compl1} the following expression
\begin{align}
	\begin{aligned}
		&\int \mathcal{D}\bm{x} \; \mu(\bm{x}) \, \Psi_{\bm{x}}(z) \, \overline{\Psi_{\bm{x}}(w)}  \\
		& \qquad = \frac{1}{2\pi^2} \frac{ [z+\gamma]^{g - s} \, [w+\gamma]^{s-g} }{ | w - \gamma|^2 }
		\int \mathcal{D}\bm{y} \, \left[\frac{w-\gamma}{z-\gamma}\right]^{y + g}.
	\end{aligned}
\end{align}
The last step is to use the relation
\begin{align}
\int \mathcal{D}\bm{y}\,
\left[\frac{v}{u}\right]^{y + g } = 2\pi^2 \, |v|^2 \, \delta^2(u-v),
\end{align}
which can be easily proved in polar coordinates $v/u = r  \mathrm{e}^{i\varphi}$. After that we obtain the answer for the completeness relation~\eqref{compl}.

\section{Reflection equations}\label{sec:refl}
\begin{figure}[t]
	\centering
	\begin{tikzpicture}[thick, line join = round, line cap = round]
		\def\l{4}
		\def\h{0.2}
		\def\d{1.5}
		\draw[line width = 0.2mm]  (0,\d) node[above]{\footnotesize$1$} -- (2, 0) -- (4, \d);
		\draw[line width = 0.2mm]  (1.8, \d) node[above]{\footnotesize$2$} -- (2.5, 0) -- (3.2, \d);
		\fill[pattern = north west lines] (0,0) rectangle (\l,-\h);
		\draw (0,0) -- (\l, 0);
	\end{tikzpicture}\hspace{0.6cm}
	\begin{tikzpicture}[thick, line join = round, line cap = round]
		\def\l{4}
		\def\h{0.2}
		\def\d{1.5}
		\node (a) at (-1,0.5*\d){$=$};
		\draw[line width = 0.2mm]  (0,\d) node[above]{\footnotesize$1$} -- (2, 0) -- (4, \d);
		\draw[line width = 0.2mm]  (0.8, \d) node[above]{\footnotesize$2$} -- (1.5, 0) -- (2.2, \d);
		\fill[pattern = north west lines] (0,0) rectangle (\l,-\h);
		\draw (0,0) -- (\l, 0);
	\end{tikzpicture}
	\caption{Reflection equation} \label{fig:reflection}
\end{figure}
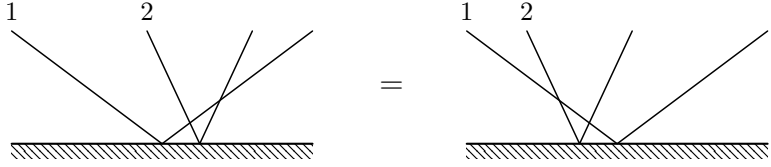

The boundary $K$-matrix
\begin{equation}
	K(u) = \begin{pmatrix}
		\imath \alpha & u - \frac{1}{2} \\[6pt]
		\gamma^2 \bigl(u - \frac{1}{2} \bigr) & \imath \alpha
	\end{pmatrix}
\end{equation}
satisfies the reflection equation
\begin{multline}\label{RKRK-1}
\bigl(\bm{1} \otimes K(v)\bigr)\,R(u+v-1)\,\bigl(K(u)\otimes \bm{1}\bigr)\, R(u-v) \\[6pt]
= R(u-v)\, \bigl(K(u)\otimes \bm{1}\bigr)\,R(u+v-1)\, \bigl(\bm{1} \otimes K(v)\bigr)
\end{multline}
with the rational $R$-matrix \eqref{R} acting in $\mathbb{C}^2 \otimes \mathbb{C}^2$
\begin{align}
	R(u) = u + P, \qquad P \, a \otimes b = b \otimes a.
\end{align}
This equation is usually depicted as in Figure \ref{fig:reflection} that we read from left to right: each reflection of lines from the ``floor'' corresponds to the $K$-matrix and each intersection of lines corresponds to the $R$-matrix. The numbers $1,2$ correspond to different matrix spaces $\mathbb{C}^2$ from the tensor product.

Defining equation
\begin{align}\label{RKRK-2}
	\begin{aligned}
		&\mathcal{K}(\bm{s},\bm{x}) \, L(u +x - 1, u - s) \, K(u) \, L(u  + s - 1, u - x) \\[6pt]
		& = L(u + s - 1, u - x) \, K(u) \, L(u + x - 1, u - s) \, \mathcal{K}(\bm{s},\bm{x})
	\end{aligned}
\end{align}
for the reflection operator
\begin{align}
\mathcal{K}(\bm{s},\bm{x}) & = [z + \gamma]^{g-s} \, [z - \gamma]^{1 - s - g}\,
[\partial_z]^{x - s} \, [z + \gamma]^{x - g} \, [z - \gamma]^{x + g - 1}
\end{align}
also can be pictured by Figure \ref{fig:reflection} with the first line ``carrying'' infinite-dimensional space of functions $\psi(z)$, on which reflection operator and elements of Lax matrices act.

It is natural to ask, whether there is a variant of reflection equation with the same reflection operator and both spaces being infinite-dimensional. The answer is given by the following identity
\begin{align}\label{RKRK-3}
	\begin{aligned}
		& \mathcal{K}_1(\bm{x}) \; \widetilde{\mathbb{R}}_{12}(\bm{x}, \bm{y}) \; \mathcal{K}_2(\bm{y}) \; {\mathbb{R}}_{12}(\bm{x},\bm{y}) \\[6pt]
		& \quad\;\; = {\mathbb{R}}_{12}(\bm{x},\bm{y}) \; \mathcal{K}_2(\bm{y}) \; \widetilde{\mathbb{R}}_{12}(\bm{x}, \bm{y}) \; \mathcal{K}_1(\bm{x}).
	\end{aligned}
\end{align}
Here the lower indices mean spaces of functions of $z_1, z_2$ and in all operators we suppressed the dependence on spin $\bm{s}$. The $\mathbb{R}$-operators are defined~as
\begin{align}
	&{\mathbb{R}}_{12}(\bm{x},\bm{y}) = \mathcal{P}_{12} \; [z_{12}]^{1 - s - x} \, [\partial_{z_1}]^{y - x} \, [z_{12}]^{s + y - 1}, \\[6pt]
	&\widetilde{\mathbb{R}}_{12}(\bm{x},\bm{y}) = \mathcal{P}_{12} \; [z_{12}]^{1 - 2s} \, [\partial_{z_2}]^{x - s} \, [\partial_{z_1}]^{y - s} \, [z_{12}]^{x + y - 1}
\end{align}
where $z_{12}\equiv z_1 - z_2$ and $\mathcal{P}_{12}$ is the permutation operator
\begin{align}
	\mathcal{P}_{12} \, \psi(z_1, z_2) = \psi(z_2, z_1).
\end{align}
Notice that for the two previous equations \eqref{RKRK-1} and \eqref{RKRK-2} we have antiholomorphic counterparts, whereas in the last one \eqref{RKRK-3} both holomorphic and antiholomorphic variables come into play.

The proof of the identity \eqref{RKRK-3} is long, but straightforward: one just uses many times chain and star-triangle relations \eqref{Chain}, \eqref{Star}, which can be written in terms of operators as
\begin{align}
	&[\partial_z]^a \, [\partial_z]^b = [\partial_z]^{a + b}, \\[6pt]
	& [z + w]^{a} \, [\partial_z]^{a + b} \, [z + w]^b = [\partial_z]^b \, [z + w]^{a + b} \, [\partial_z]^a.
\end{align}
Now let us explain how the identity \eqref{RKRK-3} can be guessed and where do such $\mathbb{R}$-operators come from. The operators ${\mathbb{R}}_{12}(\bm{x},\bm{y})$ and $\widetilde{\mathbb{R}}_{12}(\bm{x},\bm{y})$ are special cases of the more general $\mathbb{R}$-operator \cite[Section~3.1]{DM09}
\begin{align}
	\widehat{\mathbb{R}}_{12}(\bm{u}_1,\bm{u}_2|\bm{v}_1, \bm{v}_2) = \mathcal{P}_{12} \; [z_{12}]^{u_2 - v_1} \, [\partial_{z_2}]^{u_1 - v_1} \, [\partial_{z_1}]^{u_2 - v_2} \, [z_{12}]^{u_1 - v_2}
\end{align}
that permutes two Lax matrices
\begin{align}\label{Rh}
	\widehat{\mathbb{R}}_{12} \, L_1(u_1, u_2) \, L_2(v_1, v_2) = L_2(v_1, v_2) \, L_1(u_1, u_2) \, \widehat{\mathbb{R}}_{12}
\end{align}
and their antiholomorphic counterparts.
Indeed, from the above definitions we have
\begin{align}
	& {\mathbb{R}}_{12}(\bm{x},\bm{y}) = \widehat{\mathbb{R}}_{12}(\bm{u} +\bm{s} - \bm{1}, \, \bm{u} - \bm{x} \, | \, \bm{u} + \bm{s} - \bm{1}, \, \bm{u} - \bm{y}), \\[6pt]
	& \widetilde{\mathbb{R}}_{12}(\bm{x},\bm{y}) = \widehat{\mathbb{R}}_{12}(\bm{u} + \bm{x} - \bm{1}, \, \bm{u} - \bm{s} \, | \, \bm{u} + \bm{s} - \bm{1}, \, \bm{u} - \bm{y})
\end{align}
where we denoted $\bm{1} = (1, 1)$. Using these operators and reflection operator satisfying~\eqref{RKRK-2} one can rearrange the product of matrices
\begin{align}
	\begin{aligned}
		&L_2(u + y - 1, u - s) \; L_1(u + x - 1, u - s) \; K(u) \\[6pt]
		& \hspace{2cm} \times L_1(u + s - 1, u - x) \; L_2(u + s - 1, u - y) \\[10pt]
		& \to \quad L_2(u + s - 1, u - y) \; L_1(u + s - 1, u - x) \; K(u) \\[6pt]
		&\hspace{2cm} \times L_1(u + x - 1, u - s) \; L_2(u + y - 1, u - s)
	\end{aligned}
\end{align}
in two different ways, which correspond to the left and right hand sides of the identity~\eqref{RKRK-3}. Hence, we guess that operators from opposite sides of~\eqref{RKRK-3} are equal.

\section*{Acknowledgements}

The work of P.~Antonenko and S.~Derkachov (Sections 3--6, Appendix A) was supported by the Russian Science Foundation under grant No. \mbox{23-11-00311}. The work of N.~Belousov (Sections 2, 7) was supported by the Russian Science Foundation under grant No. 24-21-00466.

\appendix

\section{Diagram technique} \label{App-DiagTech}

\begin{figure}[h]
	\begin{minipage}{\textwidth}
		\centering\includegraphics[scale=0.6]{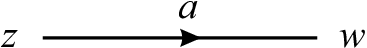}
		\caption{Diagrammatic representation of $[z-w]^{-a}$}
		\label{propag}
	\end{minipage}
\end{figure}

\noindent In the paper functions and scalar products are represented by diagrams. They consist of directed lines (Figure \ref{propag}) that correspond to the function
\begin{equation}
	\frac{1}{[z-w]^a}\equiv\frac{1}{(z-w)^a (\bar{z}-\bar{w})^{\bar a}}=
	\frac{(\bar{z}-\bar{w})^{a-\bar a}}{|z-w|^{2a}},
\end{equation}
where we always assume $a-\bar a \in \mathbb{Z}$. The flip of the arrow gives an additional sign factor
\begin{equation}
	\frac{1}{[z-w]^a}=
	\frac{(-1)^{a-\bar a}}{[w-z]^{a}}= \frac{[-1]^{a}}{[w-z]^{a}}.
\end{equation}

\begin{figure}[t]
	\centerline{\includegraphics[scale=0.7]{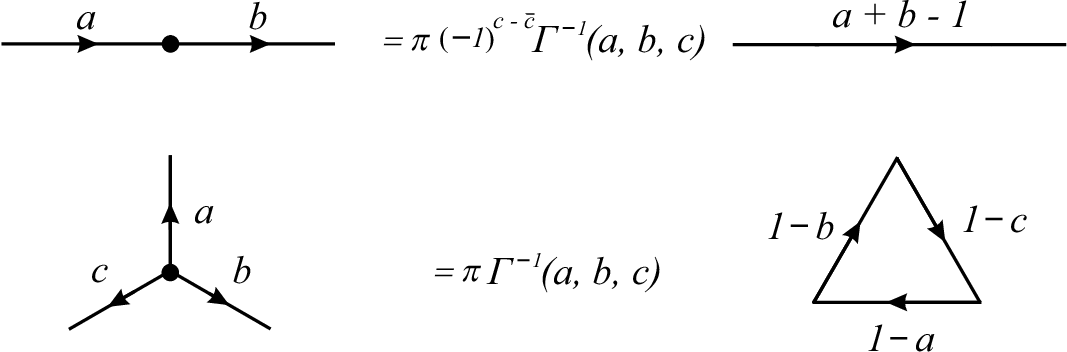}}
	\caption{The chain and star-triangle relations, $a+b+c=2$}
	\label{Rules}
\end{figure}

There are several useful integral identities, which correspond to transformations of diagrams. The proofs can be found in \cite[Appendix A]{DKM}. The first identity is the \textit{chain relation} depicted in Figure \ref{Rules}
\begin{equation}\label{Chain}
	\int \mathrm{d}^2 w \, \frac{1}{[z_1-w]^a \, [w-z_2]^{b}}=
	\frac{\pi \, [-1]^c}{\mathbf{\Gamma}(a,b,c)}
	\frac{1}{[z_1-z_2]^{a+b-1}},
\end{equation}
where $c=2-a-b,\ \bar c=2-\bar a-\bar b$. The bold vertex corresponds to the integration over $\mathbb{C}$. From the right we have gamma function of the complex field~\cite[Section~1.4]{GGR},~\cite[Section 1.3]{N}
\begin{equation}\label{gamma-def}
	\bm{\Gamma}(a) = \frac{\Gamma(a)}{\Gamma(1 - \bar{a})}
\end{equation}
and we denote its products as
\begin{align}
	\bm{\Gamma}(a,b) = \bm{\Gamma}(a)  \bm{\Gamma}(b).
\end{align}
Notice that $\bm{\Gamma}(a)$ depends on two parameters $(a, \bar{a}) \in \mathbb{C}^2$ such that $a - \bar{a} \in \mathbb{Z}$, but for brevity we display only the first one. Moreover, for $\rho \in \mathbb{R}$ we write
\begin{align}
	\bm{\Gamma}(a + \rho) \equiv \frac{\Gamma(a + \rho)}{\Gamma(1 - \bar{a} - \rho)}.
\end{align}
From the well-known properties of the ordinary gamma function it is easy to prove the following relations
\begin{align}\label{gamma-diff}
	& \bm{\Gamma}(a + 1) = - a \bar{a} \, \bm{\Gamma}(a), \\[6pt] \label{gamma-swap}
	& \bm{\Gamma}(a) = [-1]^{a} \, \bm{\Gamma}(\bar{a}), \\[6pt] \label{gamma-refl}
	& \bm{\Gamma}(a) \, \bm{\Gamma}(1 - a) = [-1]^a.
\end{align}
The second diagram identity is the \textit{star-triangle relation} shown in Figure~\ref{Rules}
\begin{align}\label{Star}
		\begin{aligned}
			&\int \mathrm{d}^2 w \, \frac{1}{[w-z_1]^a[w-z_2]^b [w-z_3]^c} \\[6pt]
			& = \frac{\pi}{\mathbf{\Gamma}(a,b,c)}
			\frac{1}{[z_1-z_2]^{1-c}[z_3-z_1]^{1-b}[z_2-z_3]^{1-a}}\,,
		\end{aligned}
\end{align}
where again $a+b+c=2$ and $\bar a+\bar b+\bar c=2$.

\begin{figure}[t]
		\begin{minipage}{\textwidth} \hspace{0.3cm}
			\centerline{\includegraphics[scale=0.5]{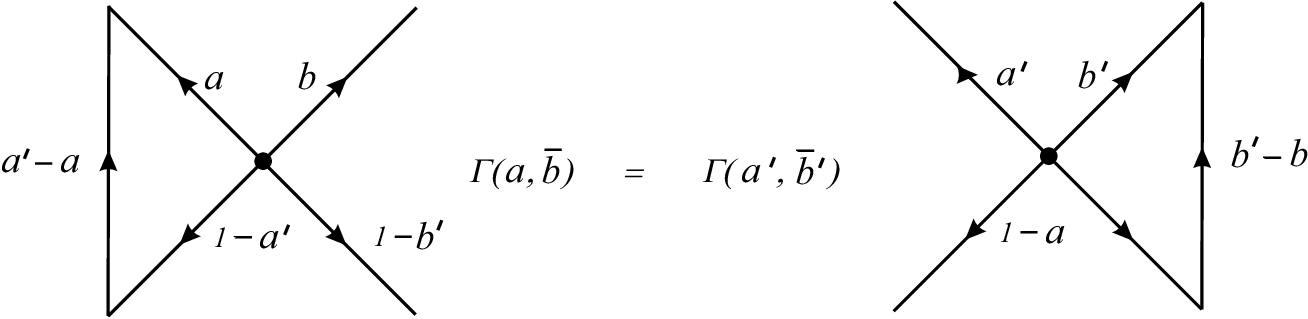}}
		\end{minipage}\vspace{0.6cm}
		\begin{minipage}{\textwidth}
			\centerline{\includegraphics[scale=0.5]{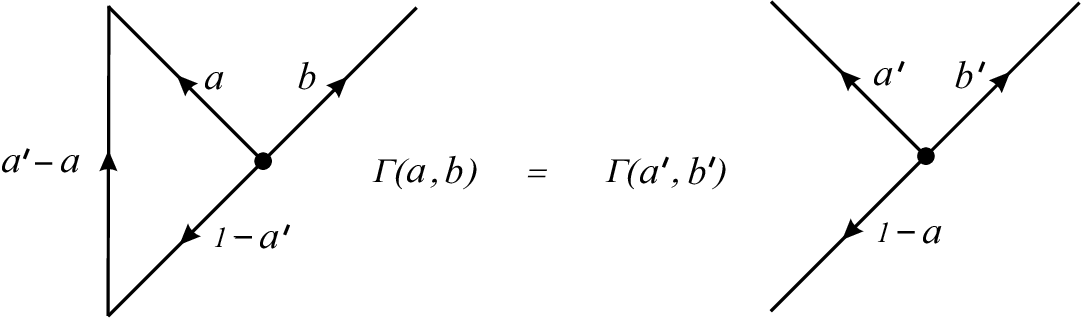}}
		\end{minipage}\vspace{0.1cm}
		\caption{Cross and reduced cross relations, $a+b=a'+b'$}
		\label{fig:Cross}
\end{figure}

From the star-triangle relation one can deduce the so-called \textit{cross relation}
\begin{align}
	\begin{aligned}
		&\frac{1}{[z_2-z_1]^{a'-a}}\int \mathrm{d}^2 w \,
		\frac{\mathbf{\Gamma}(a,\bar b)
		}{[w-z_1]^a [w-z_2]^{1-a'} \, [w-z_3]^b \, [w-z_4]^{1-b'}} \\[6pt]
		&= \frac{1}{[z_4-z_3]^{b-b'}}
		\int \mathrm{d}^2w \,
		\frac{\mathbf{\Gamma}(a',\bar b')}{[w-z_1]^{a'} \, [w-z_2]^{1-a} \,
			[w-z_3]^{b'} \, [w-z_4]^{1-b}},
	\end{aligned}
	\end{align}
depicted in Figure \ref{fig:Cross}, where we assume $a+b=a'+b'$, $\bar{a}+\bar{b}=\bar{a}'+\bar{b}'$. Its limiting case gives the \textit{reduced cross relation}
\begin{align}\label{Cross_reduced}
	\begin{aligned}
		\frac{1}{[z_2-z_1]^{a'-a}} &\int \mathrm{d}^2 w \,
		\frac{\mathbf{\Gamma}(a,b)
		}{[w-z_1]^a \, [w-z_2]^{1-a'} \,
			[w-z_3]^b} \\[6pt]
		= & \int \mathrm{d}^2w \,
		\frac{\mathbf{\Gamma}(a',b')}{[w-z_1]^{a'} \, [w-z_2]^{1-a} \,
			[w-z_3]^{b'}},
	\end{aligned}
	\end{align}
	where again $a+b=a'+b'$, $\bar{a}+\bar{b}=\bar{a}'+\bar{b}'$.
	
\begin{thebibliography}{99}
	
	\bibitem[ABDK]{ABDK} P. Antonenko, N. Belousov, S. Derkachov, S. Khoroshkin, \textit{Reflection operator and hypergeometry I: $SL(2, \mathbb{R})$ spin chain}, to appear (2024).
	
	\bibitem[C]{Ch} I. V. Cherednik, \textit{Factorizing Particles on a Half Line and Root Systems}, \href{https://doi.org/10.1007/BF01038545}{Theor. Math. Phys.} \textbf{61} (1984) 977--983.
	
	\bibitem[CDKK]{CDKK} D. Chicherin, S. Derkachov, D. Karakhanyan, R. Kirschner, \textit{Baxter operators for arbitrary spin}, \href{https://doi.org/10.1016/j.nuclphysb.2011.08.029}{Nuclear Physics B} \textbf{854}:2 (2012) 393--432, \href{https://doi.org/10.48550/arXiv.1106.4991}{\texttt{arXiv:1106.4991 [hep-th]}}.
	
	\bibitem[DKM]{DKM} S. E. Derkachov, G. P. Korchemsky, A. N. Manashov,
	\textit{Noncompact Heisenberg spin magnets from high-energy QCD: I. Baxter Q-operator and separation of variables}, \href{https://doi.org/10.1016/S0550-3213(01)00457-6}{Nuclear Physics B} \textbf{617} (2001) 375--440, \href{https://doi.org/10.48550/arXiv.hep-th/0107193}{\texttt{arXiv:hep-th/0107193}}.
	
	\bibitem[DKO]{DKO}
	S. Derkachov, V. Kazakov, E. Olivucci, \textit{Basso-Dixon correlators in two-dimensional fishnet CFT}, \href{https://doi.org/10.1007/JHEP04(2019)032}{Journal of High Energy Physics} \textbf{2019}, 32 (2019), \href{https://doi.org/10.48550/arXiv.1811.10623}{\texttt{arXiv:1811.10623 [hep-th]}}.

	\bibitem[DM1]{DM09}
	S. E. Derkachev, A. N. Manashov, \textit{General solution of the Yang–Baxter equation with symmetry group $\mathrm{SL}(n,\mathbb{C})$}, \href{https://www.mathnet.ru/rus/aa1145}{Algebra i Analiz} \textbf{21}:4 (2009) 1--94, \href{https://doi.org/10.1090/S1061-0022-2010-01106-3}{St. Petersburg Math. J.} \textbf{21}:4 (2010) 513--577.

	\bibitem[DM2]{DM14} S. E. Derkachov, A. N. Manashov, \textit{Iterative construction of eigenfunctions of the monodromy matrix for $SL(2,\mathbb{C})$ magnet}, \href{https://doi.org/10.1088/1751-8113/47/30/305204}{J. Phys. A: Math. Theor.} \textbf{47}:30 (2014), \href{https://doi.org/10.48550/arXiv.1401.7477}{\texttt{arXiv:1401.7477 [math-ph]}}.

	\bibitem[DM3]{DM20}
	S. E. Derkachov, A. N. Manashov, \textit{On Complex Gamma-Function Integrals}, \href{https://doi.org/10.3842/SIGMA.2020.003}{SIGMA} \textbf{16} (2020) 003, \href{https://doi.org/10.48550/arXiv.1908.01530}{\texttt{arXiv:1908.01530 [math-ph]}}.

	\bibitem[DMV]{DMV} S. Derkachov, A. Manashov, P. Valinevich,
	\textit{$\mathrm{SL}(2,\mathbb{C})$ Gustafson Integrals}, \href{https://doi.org/10.3842/SIGMA.2018.030}{SIGMA} \textbf{14} (2018) 030, \href{https://doi.org/10.48550/arXiv.1711.07822}{\texttt{arXiv:1711.07822 [math-ph]}}.

	\bibitem[F]{F} L. D. Faddeev, \textit{How algebraic Bethe ansatz works for integrable model}, Quantum Symmetries/Symetries Quantiques: Proceedings of the Les Houches Summer School \textbf{64} (1998) 149--211, \href{https://doi.org/10.48550/arXiv.hep-th/9605187}{\texttt{arXiv:hep-th/9605187}}.

	\bibitem[FK]{FK} L. D. Faddeev, G. P. Korchemsky, \textit{High energy QCD as a completely integrable model}, \href{https://doi.org/10.1016/0370-2693(94)01363-H}{Physics Letters B} \textbf{342} (1995) 311--322, \href{https://doi.org/10.48550/arXiv.hep-th/9404173}{\texttt{arXiv:hep-th/9404173}}.
	
	\bibitem[GGR]{GGR}
	I. M. Gelfand, M. I. Graev, V. S. Retakh, \textit{Hypergeometric functions over an arbitrary field}, \href{https://doi.org/10.1070/RM2004v059n05ABEH000771}{Russian Mathematical Surveys} \textbf{59}:5 (2004) 831--905.
	
	\bibitem[GGV]{GGV} I. M. Gelfand, M. I. Graev, N. Ya. Vilenkin, \textit{Generalized functions. Vol. 5. Integral geometry and representation theory}, \href{http://dx.doi.org/10.1090/chel/381}{AMS Chelsea Publishing} (1966).
	
	\bibitem[KS]{KS} P. P. Kulish, E. K. Sklyanin, \textit{Algebraic structures related to the reflection equations}, \href{https://doi.org/10.1088/0305-4470/25/22/022}{J. Phys. A} \textbf{25} (1992) 5963--5976, \href{https://doi.org/10.48550/arXiv.hep-th/9209054}{\texttt{arXiv:hep-th/9209054}}.
	
	\bibitem[L1]{L93} L. N. Lipatov, \textit{High energy asymptotics of multi-colour QCD and two-dimensional conformal field theories}, \href{https://doi.org/10.1016/0370-2693(93)90951-D}{Physics Letters B} \textbf{309}: 3--4 (1993) 394--396.

	\bibitem[L2]{L94} L. N. Lipatov, \textit{High-energy asymptotics of multicolor QCD and exactly solvable lattice models}, JETP Lett. \textbf{59} (1994) 596--599, \href{https://doi.org/10.48550/arXiv.hep-th/9311037}{\texttt{arXiv:hep-th/9311037}}.
	
	\bibitem[M]{M} A. N. Manashov, \textit{Unitarity of the SoV Transform for $\mathrm{SL}(2,\mathbb{C})$ Spin Chains}, \href{https://doi.org/10.3842/SIGMA.2023.086}{SIGMA} \textbf{19} (2023) 086, \href{https://doi.org/10.48550/arXiv.2303.11461}{\texttt{arXiv:2303.11461 [math-ph]}}.
	
	\bibitem[MN]{MN} V. F. Molchanov, Yu. A. Neretin, \textit{A pair of commuting hypergeometric operators on the complex plane and bispectrality}, \href{https://doi.org/10.4171/JST/349}{Journal of Spectral Theory} \textbf{11}:2 (2021) 509--586, \href{https://doi.org/10.48550/arXiv.1812.06766}{\texttt{arXiv:1812.06766 [math.FA]}}.
	
	\bibitem[N]{N} Y.A. Neretin, \textit{Barnes--Ismagilov Integrals and Hypergeometric Functions of the Complex Field}, \href{https://doi.org/10.3842/SIGMA.2020.072}{SIGMA} \textbf{16} (2020) 072, \href{https://doi.org/10.48550/arXiv.1910.10686}{\texttt{arXiv:1910.10686 [math.CA]}}.

	\bibitem[S1]{S1} E. K. Sklyanin, \textit{The Quantum Toda Chain}, \href{https://doi.org/10.1007/3-540-15213-X_80}{Lect. Notes Phys.} \textbf{226} (1985) 196--233.

	\bibitem[S2]{S2} E. K. Sklyanin, \textit{Separation of Variables: New Trends}, \href{https://doi.org/10.1143/PTPS.118.35}{Prog. Theor. Phys. Suppl.} \textbf{118} (1995) 35--60, \href{https://doi.org/10.48550/arXiv.solv-int/9504001}{\texttt{arXiv:solv-int/9504001}}.

	\bibitem[S3]{S}
	E. K. Sklyanin, \textit{Boundary Conditions for Integrable Quantum Systems}, \href{https://doi.org/10.1088/0305-4470/21/10/015}{J. Phys. A} \textbf{21} (1988) 2375--2389.
	
	\bibitem[SS]{SS} G. A. Sarkissian, V. P. Spiridonov, \textit{The endless beta integrals}, \href{https://doi.org/10.3842/SIGMA.2020.074}{SIGMA} \textbf{16} (2020) 074, \href{https://doi.org/10.48550/arXiv.2005.01059}{\texttt{arXiv:2005.01059 [math-ph]}}.

\end{thebibliography}
\end{document}